\documentclass[12pt]{article}
\usepackage[margin=2.54cm]{geometry} 
\usepackage[utf8]{inputenc}
\usepackage{booktabs}
\usepackage[hidelinks,breaklinks]{hyperref} %
\usepackage[natbibapa]{apacite} 
\usepackage{graphicx}
\usepackage{float}
\usepackage{setspace}
\usepackage[hyphenbreaks]{breakurl}
\graphicspath{{figure/}}

\usepackage{fancyhdr}
\usepackage{threeparttable}

\usepackage{authblk}
\author[1, 2]{\small Scott A.\ Hale}
\author[1]{\small Adriano Belisario}
\author[2]{\small Ahmed Mostafa}
\author[3, 4]{\small Chico Camargo}

\affil[1]{\footnotesize Oxford Internet Institute, University of Oxford}
\affil[2]{\footnotesize Meedan}
\affil[3]{\footnotesize Department of Computer Science, University of Exeter}
\affil[4]{\footnotesize Ewha Womans University}

\title{Analyzing Misinformation Claims During the 2022 Brazilian General Election on WhatsApp, Twitter, and Kwai}
\date{}

\begin{document}

\maketitle

\begin{abstract}
\noindent This study analyzes misinformation from WhatsApp, Twitter, and Kwai during the 2022 Brazilian general election. Given the democratic importance of accurate information during elections, multiple fact-checking organizations collaborated to identify and respond to misinformation via WhatsApp tiplines and power a fact-checking feature within a chatbot operated by Brazil's election authority, the TSE. WhatsApp is installed on over 99\% of smartphones in Brazil, and the TSE chatbot was used by millions of citizens in the run-up to the elections. During the same period, we collected social media data from Twitter (now X) and Kwai (a popular video-sharing app similar to TikTok). 
Using the WhatsApp, Kwai, and Twitter data along with fact-checks from three Brazilian fact-checking organizations, we find unique claims on each platform. Even when the same claims are present on different platforms, they often differ in format, detail, length, or other characteristics. Our research highlights the limitations of current claim matching algorithms to match claims across  platforms with such differences and identifies areas for further algorithmic development. Finally, we 
perform a descriptive analysis examining the formats (image, video, audio, text) and content themes of popular misinformation claims.
\\\textit{Keywords}: elections, Brazil, social media, misinformation
\end{abstract}

\section{Introduction}

User-generated content platforms including Twitter (now X), WhatsApp, Kwai, and others are shaped by their users and designs. Adoption of platforms is uneven across demographics \citep{ruths2014social,jungherr2018normalizing}, and this can easily lead to unique content on each platform. For example, each edition of Wikipedia has thousands of unique articles not found on other editions \citep{hecht2010tower}, and clearly the content on a platform like Truth Social run by a company founded by former US President Donald Trump will differ from a platform like Facebook. The unique designs of platforms also signal different affordances \citep{gibson2014theory}, which could lead to unique content. For example, it's clear that Instagram favors visual content, Facebook allows longer form text, and Twitter has lots of shorter-form content.

At the same time, many people use multiple platforms in their everyday lives \citep{blank2020oxis}, and these users are embedded in various environments and social contexts with a range of interactions with other individuals both online and offline \citep{lamb2003informational}. This use of multiple platforms provides a possible mechanism for the same content to spread across platforms. In some settings, for example, it's common to find screenshots of Twitter posts on Instagram \citep{disinfotable2022power}.

Installed on over 99\% of smartphones in Brazil, WhatsApp is by far the most used platform in Brazil \citep{whatsapp99}, with 57\% of Brazilians using it for news consumption or a source of political information \citep{newman_digital_2022,anita2019circulaccao}. Yet, very little is known about the content spreading on this platform due to its end-to-end encryption. In India, misinformation on WhatsApp has been linked to violence and death \citep{banaji2019whatsapp}. Identifying and responding to misinformation is essential for transparent elections and well-functioning democracies: as Nobel Laureate Maria \citet{ressanobel} articulated it: 
\begin{quote}
    Without facts, you can’t have truth. Without truth, you can’t have trust. Without trust, we have no shared reality, no democracy, and it becomes impossible to deal with our world’s existential problems\ldots %
\end{quote}

In this paper we investigate two research questions:\\

\noindent \textbf{RQ1} To what extent does information overlap between fact-checker WhatsApp tiplines and the national TSE WhatsApp bot during the 2022 Brazilian elections?\\

\noindent \textbf{RQ2} What overlap is there between misinformation sent to fact-checker WhatsApp tiplines with content found on Twitter and Kwai during the 2022 Brazilian elections?\\

On the one hand, affordance theory suggests each of our datasets will have unique characteristics and contain unique misinformation based on the perceived affordances of each platform or app. On the other hand, the fact that many people use multiple platforms in their everyday lives points provides a mechanism for content to spread from one platform to another suggesting a higher degree of overlap. The answers to these questions have important practical implications for misinformation response: if misinformation is relatively homogeneous across different platforms then fact-checking organizations can monitor whatever platforms are most easily accessible. In contrast, if misinformation is relatively unique per platform there is a greater need for gathering content in platform-specific ways. This research was undertaken prior to dramatic changes in data access to Twitter (now X) in Spring 2023, but such restrictions on data access make these research questions all the more pressing.

To the best of our knowledge, this paper is the first large-scale analysis of misinformation using data from Kwai, Twitter, and WhatsApp. In particular, we're unaware of any prior peer-reviewed scholarship on Kwai, a popular short-video platform that is similar in some respects to TikTok. Similarly, our crowdsourcing approach to gathering WhatsApp data differs from much existing scholarship which has focused on analyzing messages collected from large WhatsApp groups found via the Internet \citep[e.g.,][]{resende_analyzing_2019,resende_misinformation_2019}. 

Our study is set during the 2022 Brazilian election. The first round of voting was held on 2 October and included elections for the houses of the Brazilian Congress, state governors, and the president. After the first round, no presidential candidate captured over 50\% of the ballots cast, and a run-off election was set for 30 October between the top two candidates: Luiz Inácio Lula da Silva (former president of Brazil, from 2003 to 2010) and Jair Bolsonaro (elected in 2018, seeking re-election). In the end, Lula da Silva was elected for his third mandate, defeating the incumbent Bolsonaro.

Overall, we find the unique characteristics of each platform (formats, text length limits, etc.) influence the specific content circulating on the platforms. These differences also complicate any large-scale, quantitative comparison, and our paper highlights areas where further algorithmic development is needed. While there is overlap across the platforms and the heavy-tailed patterns often observed in collective human behavior \citep{margetts2015political} can be observed here, we also find clear differences between platforms. There is a clear separation, for instance, between the videos we capture from WhatsApp and Kwai. The nature of the interface also affects use: the WhatsApp tiplines operated by fact-checkers capture many, long forwarded messages whereas people submitted shorter, search-query-like messages to the bot operated by Brazil's election authority, the TSE. 

The remainder of this paper discusses the context of the Brazilian elections and social media as well as related work in Section \ref{sec:literature}. We describe our data sources and methods (Section \ref{sec:data-methods}), present results (Section \ref{results}), and discuss wider implications (Sections \ref{discussion} \& \ref{sec:conclusion}).

\section{Related work} \label{sec:literature}
We first introduce the broad context of 2022 Brazilian elections and social media use, giving details about the role of WhatsApp, Kwai, and Twitter and their interactions with the electoral process. We then discuss misinformation online and its interaction with affordance theory, imagined audiences, and the embedded use of multiple platforms in everyday life.

\subsection{Social media and the 2022 Brazilian general election}

\label{sec:context}

Social media platforms have been a critical part of contemporary Brazilian politics, serving as both a catalyst and a conduit for political discourse, activism, and misinformation \citep{resende_misinformation_2019, reis_can_2020, reis_dataset_2020, dwoskin_come_2023}. Digital platforms provide an alternative space where politicians can connect directly with the electorate, disseminate their messages, and set agendas. Beyond that, these platforms have enabled grassroots movements to amplify their voices, exemplified by various social campaigns and political protests that have found traction online \citep{margetts2015political}. However, this democratization of voice and space has also led to the propagation of misinformation and the polarization of public opinion \citep{kischinhevsky_whatsapp_2020}. In Brazil, this dynamic is particularly noteworthy, as these platforms are not just secondary channels but often primary interfaces where political ideologies are formed, debated, and propagated \citep{batista_pereira_fake_2022}. Whether it's the spread of election-related rumors, discussions about corruption, or the mobilization for or against government policies, social media has become a critical battleground that reflects and, at times, amplifies the complexities of Brazilian politics \citep{machado_study_2019}.

Brazil is the largest country in Latin America, with approximately 217 million people, and has significant Internet penetration. Of the Brazilian population aged 10 or older, 81\%---148 million individuals---were Internet users in 2021 \citep{ceticbr_survey_2021}. While the raw number of Internet users in Brazil is high, it is important to note that most of the population accesses the Internet exclusively through mobile phones and mobile networks subject to data caps \citep{ceticbr_survey_2021}.

Instant message applications, and WhatsApp in particular, are considerably more popular than social networking websites: 93\% of all Brazilian Internet users exchange instant messages and 81\% access social networking sites \citep{ceticbr_survey_2021}. This context is partially derived from, and reinforced by, zero-rating plans that allow the free use of certain applications.\footnote{There are various studies on the zero-rating effects on net neutrality and the Brazilian Civil Rights Framework for the Internet \citep{belli_governance_2018,belli_net_2017,pereira_compatibility_2019}, as well as the spread of disinformation in Brazil \citep{tomaz_brazilian_2020,evangelista_whatsapp_2019}.}
Each mobile operator decides the applications they want to offer for free, but WhatsApp is an omnipresent choice. 

Table~\ref{tab:brazil-socmedia} reproduces WhatsApp, Kwai, and Twitter metrics from recent surveys about social media usage in Brazil. Dashes represent responses not applicable or not reported. Despite having different methodologies, limitations, and target audiences, the overall picture is clear: WhatsApp is by far the most popular application, but both Twitter and Kwai also have considerable user bases in the country.

\begin{table}[tbh]
\centering
\caption{Online surveys on social media usage in Brazil}
\label{tab:brazil-socmedia}
\begin{tabular}{@{}lrrr@{}}
\toprule
\textbf{Prompt or topic} & \multicolumn{1}{l}{\textbf{WhatsApp}} & \multicolumn{1}{l}{\textbf{Kwai}} & \multicolumn{1}{l}{\textbf{Twitter}} \\ \midrule
\textit{\begin{tabular}[c]{@{}l@{}}Do you have Facebook, Instagram, \\ Kwai, LinkedIn, TikTok, Twitter \\ installed on your smartphone? \\ \citep[~p.10]{paiva_panorama_2022}\end{tabular}} & - & 32\% & 37\% \\ \midrule
\textit{\begin{tabular}[c]{@{}l@{}}What application do you\\ spend more time throughout the day? \\ \citep[~p.7]{paiva_panorama_2022}\end{tabular}} & 30\% & 3\% & - \\ \midrule
\textit{Most-used social media platforms} \\ \citep{kemp_digital_2022} & 96\% & 62\% & 48\% \\ \midrule
\textit{Most-used digital communication platform} \\ \citep{gava_estudo_2022} & 94\% & 51\% & 51\% \\ \bottomrule
\end{tabular}
\end{table}

The exchange of private or group messages on these platforms plays a key role in shaping public opinion in Brazil. Social networks and WhatsApp groups have been consistently used as information sources, and traditional press sources face distrust. The proportion of respondents who say that ``they avoid the news, sometimes or often, has doubled in Brazil (54\%) and the UK (46\%) since 2017---and also increased in all other markets'' \citep[~p.13]{newman_digital_2022}. Traditional media are considered politically biased by many Brazilians \citep[~p.41]{newman_digital_2022}.

Much like many recent democratic elections, the run-up to the Brazilian general elections was rife with claims of illegitimacy of the electoral process, as well as misleading information regarding multiple aspects of Brazilian politics, such as the potential role of the Brazil's Supreme Electoral Court (TSE) \citep{tarouco_brazilian_2023,rossini_explaining_2023}. Concerned with the spread of online misinformation during the electoral period, Brazil's TSE signed agreements with WhatsApp, Twitter, Kwai and other major platforms to fight misleading content about the electoral processes of the 2022 general elections. Below, we contextualize the presence of each social network analyzed and present relevant facts about their involvement with the Brazilian elections.

\paragraph{WhatsApp} %
is Brazil's most popular application and the second most popular for news consumption \citep[~p.117]{newman_digital_2022}. 
The TSE initiated a partnership with WhatsApp in the local elections in 2020 and renewed it in 2022. Among other actions, WhatsApp offered support to develop a chatbot for the TSE \citep{tse_eleicoes_2022}. This chatbot distributed election information and had a feature via which questions and content could be submitted to fact-checkers. The WhatsApp data we analyse comes, in part, from anonymous questions submitted via that fact-checking feature. 
 
\paragraph{Kwai} is owned by Kuaishou Technology, a Chinese company which operates multiple short-video platforms and is usually cited as the main rival to Tiktok. The company has different regional brands: it is known as Snack Video in South Asia and Kwai in Latina America. Kuaishou's platforms now have more than 1 billion users globally. The company opened its Brazilian branch in 2019 and claimed to have 45 million monthly active users in Brazil during the 2022 general election, which corresponds to roughly 20\% of the country's population; most of them are located in less economically privileged regions in Brazil such as the country's North and Northeast \citep{deck_tiktoks_2022}. In its partnership with the TSE for the 2022's elections \citep{oliveira_memorando_2022}, Kwai agreed to create special content about Brazilian elections, promote outreach information provided by TSE, and provide training for the TSE staff. 

\paragraph{Twitter} (now X) is a micro-blogging service. It has approximately 450 million monthly active users, and it opened an office in Brazil in 2013. At the time of the elections, the user base in the country was estimated at 19.05 million \citep{kemp_digital_2022}. Although not as popular as WhatsApp (and maybe even Kwai), Twitter data presented a number of important features. %
The platform is popular among key actors with influence on public opinion including politicians, journalists, researchers, activists, and company executives. Twitter was intensively used for political debate in the 2018 general elections, and the role of bots or automated accounts in misinformation networks has been documented by researchers \citep{recuero_fraudenasurnas_2020}. In its agreement with the TSE for the 2022 elections, Twitter agreed to activate search prompts and warnings with official information on top of content related to elections, create special content, and provide diligent and fast follow-up on misinformation denounciatons made by the TSE \citep{oliveira_memorando_2021}.

\subsection{Misinformation online}
Scholarship about `misinformation' online appears only a few years after the first graphical web browser: \citet{hernon1995disinformation} defines disinformation as a `deliberate attempt to deceive or mislead' and misinformation as `an honest mistake' (p. 134). Similar to other scholars, we use the term `misinformation' as an overarching umbrella term for any misleading content regardless of the author's intentions \citep[see, e.g.,][]{pantazi2021social}.\footnote{We include studies of `fake news' but refrain from using the term here given its imprecision and politicization \citep{brummette2018politicization}.}

As Internet use has grown so too has scholarship on misinformation online \citep{ha_mapping_2021}, but misinformation is not a problem exclusive to the Internet \citep{altay2023misinformation}. There are a variety of psychological processes at play in the spread of misinformation including excess gullibility and excess vigilance \citep{pantazi2021social}. At the same time, motivated reasoning or directional motives---that is, the desire to arrive at a specific conclusion---also play a role, specifically in political misinformation \citep{jerit2020political}. 

There is a long scholarship on how people perform online to their `imagined audiences'---the mental conceptualization one has of the people who will see their posts \citep{litt2012imagined}---and people spread misinformation for a variety of reasons including entertainment and sarcasm  \citep{metzger2021from} and, for political misinformation, because of hostile feelings toward a different political party \citep{osmundsen2021partisan}.

As people imagine their audiences differently on each user-generated content platform, it's probably that the content they produce will also differ \citep{litt2012imagined}. For example, users likely post differently on a `professional' networking site such as LinkedIn than they do on a dating website or on one for socializing with friends. In turn, we might expect different misinformation to be present on different websites. 

Similarly, the design of social media platforms can also influence the spread of misinformation as design can influence how much people think about accuracy of the content they are sharing \citep{pennycook2021shifting}. More broadly, the design of user interfaces influences what actions users perceive are available for them and the intended uses of a platform \citep{gibson2014theory, norman2013design}. 

At the same time, however, users often access multiple platforms and are embedded across a range of environments and social contexts \citep{lamb2003informational}. This provides a mechanism via which similar content could be shared across multiple user-generated content platforms.

While most studies focus on one-platform---often Twitter, which has generally been over-researched to the detriment of other platforms \citep{cihon2016twitter}---few studies compare misinformation content across multiple languages.
Content on WhatsApp has generally been studied through joining `public groups,' which are groups that can be joined via link on the Internet \citep{garimella2018whatsapp}, or by asking administrators of groups to consent \citep{chauchard2022what}. 
As explained further in the following sections, our WhatsApp data comes from misinformation `tiplines' and the fact-checking feature of the TSE WhatsApp bot. Tiplines are accounts to send potential misinformation to fact-checkers and discover fact-checks \citep{johansson2022combat}. While less common than other approaches, tiplines have a number of advantages in terms of preserving privacy and collecting content actually viewed by and concerning to people \citep{kazemi2022tiplines}. 

In this work, we examine two related research questions. First, we ask to what extent information overlaps between fact-checker WhatsApp tiplines and the national TSE bot during the 2022 Brazilian elections? (RQ1). These data all come from WhatsApp but are solicited in different ways. Each fact-checking organization runs its own tipline, and each has its own audience and unique organizational environment \citep{thales2022rise}. Examining the overlap between the content sent to different fact-checkers or to the TSE bot will help us understand how homogenous content is across different tiplines. If the content is quite diverse it will suggest that each tipline captures only a small proportion of possible misinformation on WhatsApp.

Our second research question asks about the overlap between misinformation sent to fact-checker WhatsApp tiplines with content found on Twitter and Kwai during the 2022 Brazilian elections (RQ2). Given the national focus on the election in Brazil and the fact that many people use multiple platforms, it's possible for similar misinformation content to exist on all three of our platforms (WhatsApp, Twitter, and Kwai). At the same time, however, the affordances of each platform and the different audiences people imagine for posts on each platform could lead to large differences. As mentioned earlier, this is also a practical question affecting how data should be collected for fact-checking and for academic research.

\section{Data and methods} \label{sec:data-methods}
This section outlines the data and methods we use to analyze text, images, and videos across WhatsApp tiplines, Kwai, and Twitter. We describe the processes for collecting and assessing data from these platforms, including automated and manual analyses. We detail the techniques used to vectorize and cluster text, video, and images, as well as the statistical tests and measurements to compare data across multiple tiplines and platforms.

\subsection{Data sources}

Table \ref{tab:data-collection} shows a summary of the data sources. Twitter and WhatsApp posts include text, images and videos, whereas Kwai posts comprise only videos and text for the video descriptions.

\begin{table}[tbh]
\centering
\caption{Data collection methods and statistics for social media platforms}
\begin{threeparttable}
\label{tab:data-collection}
\begin{tabular}{llrr@{}l}
\toprule
\textbf{Method} & \textbf{Date range (inclusive)} & \textbf{N. of posts} & \textbf{Users} \\\midrule
Twitter & 09/20/2022--11/10/2022\textsuperscript{\dag} & 53,831,265 & 4,217,513 \\
Kwai & 10/08/2022--11/26/2022 & 23,737 & 13,068 \\
WhatsApp (tiplines) & 09/01/2022--11/15/2022 & 49,422 & 14,959\\
WhatsApp (TSE bot) & 09/01/2022--11/15/2022 & 223,621 & unknown &\textsuperscript{\ddag} \\ \bottomrule
\end{tabular}
    \begin{tablenotes}[flushleft]
        \small 
        \item[\textsuperscript{\dag}] Twitter data collected 09/20/2022--10/06/2022 and 10/21/2022--11/10/2022.
        \item[\textsuperscript{\ddag}] The TSE bot was anonymous and hence the number of unique users is unknown.
    \end{tablenotes}
\end{threeparttable}
\end{table}

\subsubsection{WhatsApp}
\label{whatsapp-data-source}

WhatsApp has been widely used in Brazil for communication on political issues, and previous studies have documented the role of WhatsApp groups and bulk messages for misinformation networks in the 2018 general elections \citep{machado_study_2019,resende_misinformation_2019}. These studies usually rely on an analysis of messages exchanged in large groups that are linked publicly on the Internet \citep{machado_study_2019,resende_analyzing_2019}. In contrast to this, we use a crowdsourcing approach working with fact-checking organizations running misinformation tiplines. Tiplines are accounts on WhatsApp to which users can send possible misinformation content and questions and in exchange receive fact-checks and trusted information \citep{johansson2022combat}. Analysis of public group and tipline data in the 2019 Indian elections found tiplines capture a significant proportion of popular content and identify that content quickly—often before it spreads in large groups \citep{kazemi2022tiplines}.

WhatsApp is end-to-end encrypted and as such there is no way to capture a representative sample of the data following through the platform. We analyze anonymized data from three fact-checking organizations operating such tiplines in Brazil during the elections. We also include anonymous content sent to these fact-checking organizations via the fact-checking feature on the chatbot operated by the TSE and heavily promoted by WhatsApp. Our data includes 49,422 submissions to fact-checker tiplines from 14,959 unique users as well as 223,621 submissions from an unknown number of TSE bot users between 1 September and 15 November 2022.
All data is anonymous: phone numbers were replaced with random ids before we received the data, and no other metadata beyond the timestamp was included in the data made available to us for analysis.

\subsubsection{Kwai}
Although Kwai does not have an official API, we found a third-party service with APIs for the platform.\footnote{\url{https://rapidapi.com/contact-cmWXEDTql/api/kwai4}} We used this service and validated it by comparing its data in real-time with the search results provided by queries run by a Kwai user in Brazil on 23 October 2022, using the same keywords. We found that the top videos suggested and their metrics were the same. 

We ran queries to this API with our location set to Brazil between 8 October and 26 November 2022. We added search keywords incrementally and found the results were sensitive to accents and capitalization; to account for this, we created variations for some of the search terms. We collected 589,878 search results and group them by user and video identification variables. This means that videos published by distinct usernames count as different posts. The search results corresponded to 35,701 unique videos that we download, cluster, and compare with WhatsApp. After preprocessing the video descriptions, as described below in Section \ref{subchap:textcluster}, we identify 15,017 video descriptions that were used to cluster textual data in Kwai.

\subsubsection{Twitter}
We capture tweets from Twitter using elevated access to its streaming API, which allowed us to avoid any rate limits. We capture tweets using a list of terms related to Brazilian politics that were initially sourced from news, Wikipedia, and social media content about the elections at the start of September 2022. After one day of content was captured, we analyzed the frequencies of unigrams, bigrams, and trigrams to identify additional terms to track. In total, 128 terms were tracked from September 20 to October 6, 2022, and from October 21 to November 10, 2022. The period in between was not tracked due to a data outage. The terms matched between 460 thousand and 2.7 million tweets per day (mean 980,829; standard deviation 552,957).

\begin{figure}
    \centering
    \includegraphics[width=\textwidth]{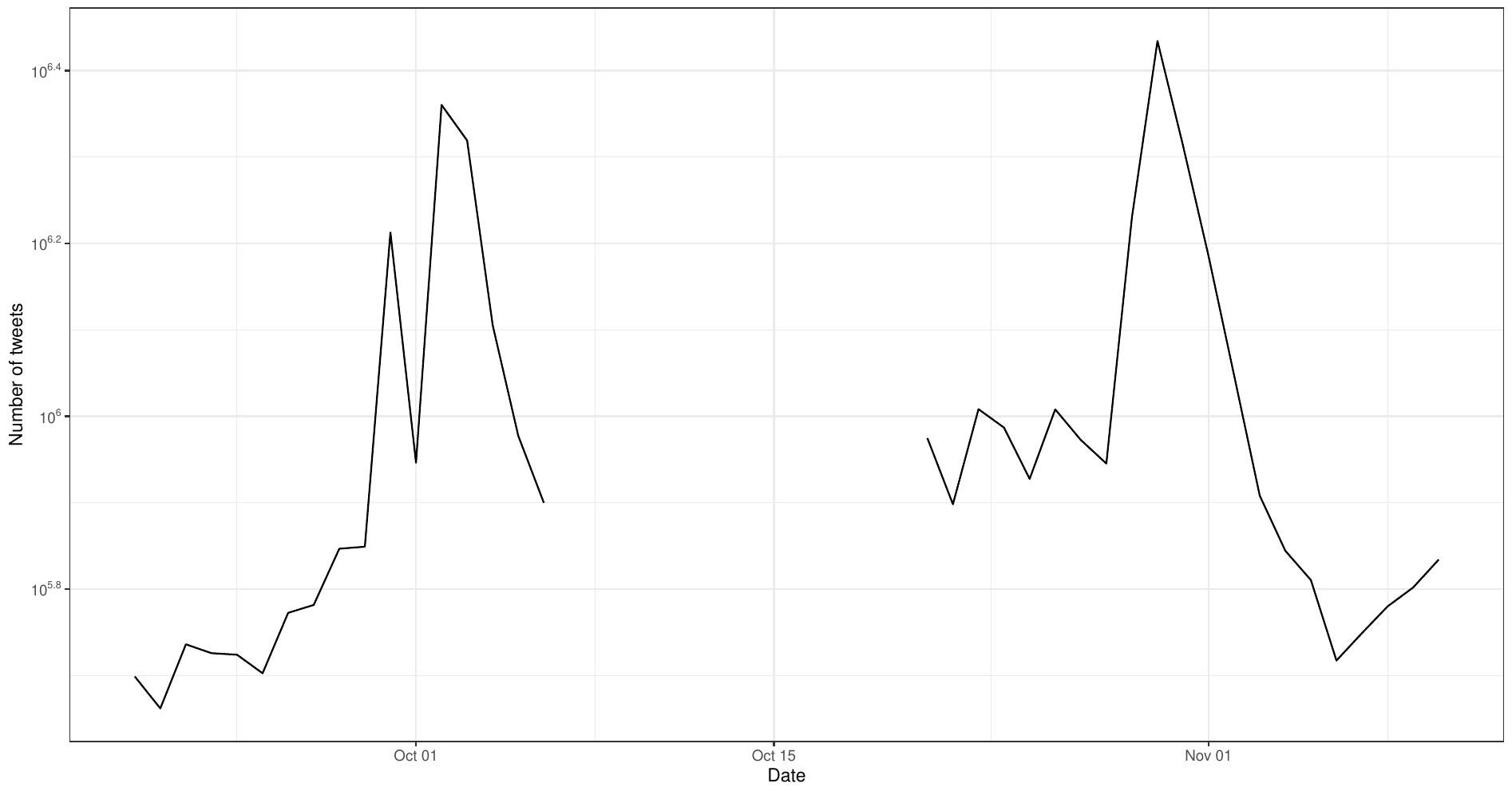}
    \caption{Volume of filtered, election-related tweets during the study period. Only original tweets in Portuguese are included (i.e., retweets are excluded).}
    \label{fig:twittervolume}
\end{figure}

We filter all tweets in a second pass and discard all retweets. We detect the language of each tweet by first removing URLs and mentions and then applying the compact language detector (CLDv3).\footnote{\url{https://github.com/google/cld3}}
Tweets not in Portuguese are discarded from our dataset. This results in a total of 53.8 million tweets from 4.2 million unique users.
We find an average of 
approximately 800,000 original tweets per day, authored by approximately 380,000 unique users each day.
The volume of tweets over time shows two clear peaks corresponding with the elections (Figure~\ref{fig:twittervolume}).

\subsection{Vectorization and clustering}
\subsubsection{Text clustering}
\label{subchap:textcluster}
We represent text content from WhatsApp, Kwai, and Twitter as dense vectors using a sentence-transformers MPNet language model trained to produce semantic sentence embeddings.\footnote{\url{https://huggingface.co/sentence-transformers/paraphrase-multilingual-mpnet-base-v2}} This model produces similar embeddings (i.e., vectors) for content with similar meanings even if the content uses different words \citep{reimers-gurevych-2019-sentence}. 

We use these embeddings directly for WhatsApp and Kwai text when comparing within each platform and across the platforms. We compare all items to each other using cosine similarity and group items that are very similar. Our goal is to only group items that are really making the same claim, and we expect many items will be in a cluster by themselves. Therefore, we use single-link hierarchical clustering, which allows us to set a similarity threshold for how similar two items need to be to group together. This contrasts with k-means clustering, where the number of clusters ($k$) is specified and items often end up in larger clusters. Based on a manual examination of our dataset and previous work on this field \citep[e.g.,][]{kazemi_claim_2021}, we use a threshold of 0.875. This approach to text clustering has been used and validated in previous misinformation research on WhatsApp \citep[e.g.,][]{kazemi_claim_2021,kazemi2021claim}.

In our initial testing, we found that capitalization, punctuation, and missing diacritics negatively affected our ability to match content across platforms. In particular, fact-checks often contained correct capitalization and diacritics, while social media content did not. In order to overcome this and make content more comparable, we took the text of each post, removed URLs, lower-cased all words, removed all punctuation, and replaced accents and other diacritics with their closest ASCII equivalent (e.g., we replaced ``á" with ``a"). 

\subsubsection{Image and video clustering}
We clustered similar images using PDQ image hashing and similar videos using TMK embeddings. Both algorithms were developed by Facebook (Meta) Research to identify similar multimedia content and are described in detail in a whitepaper available online.\footnote{\url{https://github.com/facebook/ThreatExchange/blob/main/hashing/hashing.pdf}} PDQ is a perceptual hashing algorithm that can identify similar pictures even if they have different file format or minor alternations. We calculate a normalized Hamming distance and use one minus this distance as a similarity measure. In line with previous research and the Meta whitepaper, we use a threshold of 0.7 for clustering images.  PDQ is also employed by \citet{reis2020whatsapp} in their study of image sharing on WhatsApp.

We used TMK (Temporal Match Kernel) embeddings to compare the videos received on WhatsApp tiplines (including those forwarded by the TSE tipline) to Kwai videos. TMK embeddings only consider the visual portion of the video and not the audio. In addition to the TMK embeddings, we use the related \texttt{tmk-clusterize} executable provided by Meta to identify clusters of videos. We use similarity thresholds of 0.7 for both the level-1 and level-2 thresholds in accordance with the whitepaper produced by Meta\footnote{We confirm that a more stringent threshold of 0.9 does not significantly change the results.}. TMK is a C++ library, but we develop and release open-source Python bindings to make it easier to calculate TMK embeddings in Python.\footnote{\url{https://github.com/meedan/tmkpy}}

\subsection{Analyses}

\subsubsection{WhatsApp tiplines}
To measure the information overlap between the three fact-checker WhatsApp tiplines (RQ1), we adopt the following metrics and tests. First, we count the frequency of messages according to their type (video, images, link, text or audio), pseudonymous author ids, and dates. We also measure the overlap in users between tiplines.

One challenge in crowdsourcing the identification of misinformation via tiplines is building an active user base that will forward dubious content to fact-checking organizations. The relationship between the size of a tipline audience and the amount of novel content is unclear: having a larger audience may simply result in multiple submissions of the same claim rather than the identification of new misinformation claims. We investigate this relationship, in three ways:
    (1) we measure the overlap in users between tiplines for different types of messages;
    (2) we analyze the cumulative distributive function (CDF) of the number of messages per user and the number of items per cluster (cluster sizes);
    and (3)  
    we randomly re-order users and consider the amount of novel content we could find with different subsets of random users.

We also compare data from the fact-checkers tipline to the messages received by the TSE. The comparison assesses the total, length and type of messages in both data sources. We examine the overlap and correlation of messages in Section \ref{tse-analysis}. 

\subsubsection{Kwai videos and descriptions}
To understand to what extent information sent to fact-checker WhatsApp tiplines exists on Kwai (RQ2), we first perform a descriptive analysis of the data collected from Kwai, describing overall characteristics and the most important semantic clusters for the text descriptions of videos that circulated in this platform during the election. Videos without descriptions were omitted for this part of the analysis. We also cluster the videos and examine the overlap between them and the videos sent to fact-checkers via WhatsApp.

\subsubsection{Twitter text}
To understand to what extent information sent to fact-checker WhatsApp tiplines exists on Twitter (RQ2), we select a representative message from the largest four clusters of WhatsApp messages sent to fact-checker tiplines. Then, we embed these top WhatsApp messages and all tweets and select any pairs with cosine similarity above 0.7.

Manually inspecting the Twitter messages revealed that the semantic matches were often noisy. The messages sent to the WhatsApp tiplines were often very long---in fact, they often exceeded the 280-character limit that Twitter imposes---and this may, in part, explain the poor quality of the matches between the two platforms. To overcome the noise in the results, we further filtered the close semantic matches using keywords and manual inspection.

\section{Results}
\label{results}

\subsection{Comparison of WhatsApp tiplines}

\subsubsection{Fact-checker tiplines}
\label{sec:fact-check}
We start by examining the 49,422 submissions made to the three Brazilian fact-checkers' WhatsApp tiplines.  
Most items submitted to the tiplines are videos (16,604) and images (11,463). There were a number of hyperlinks (11,363) and text messages (8,613) with a smaller number of audio messages (1,379). We exclude 1,914 short text messages of less than 5 characters for the remainder of the analysis.

\begin{figure}[bt]
    \centering
    \includegraphics[width=0.5\textwidth]{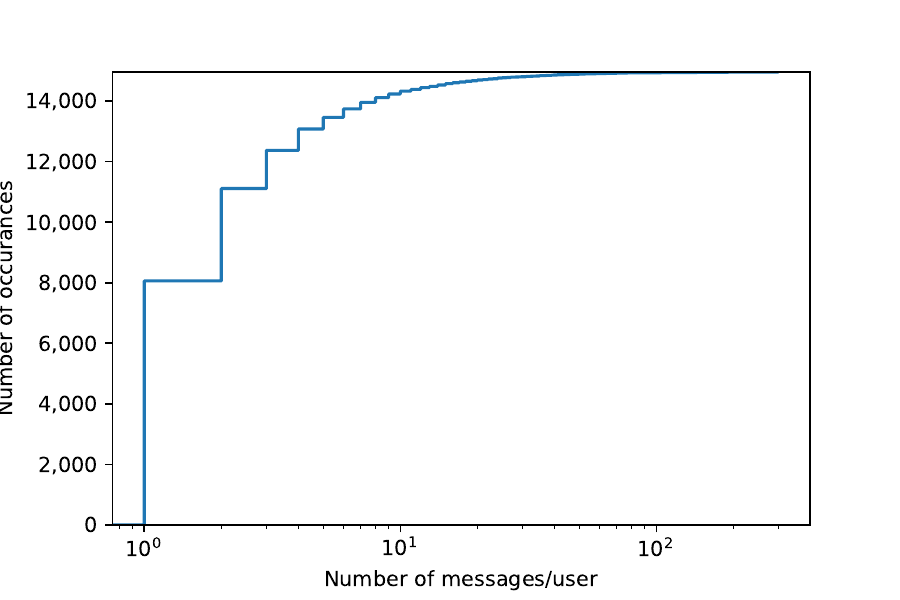}%
    \includegraphics[width=0.5\textwidth]{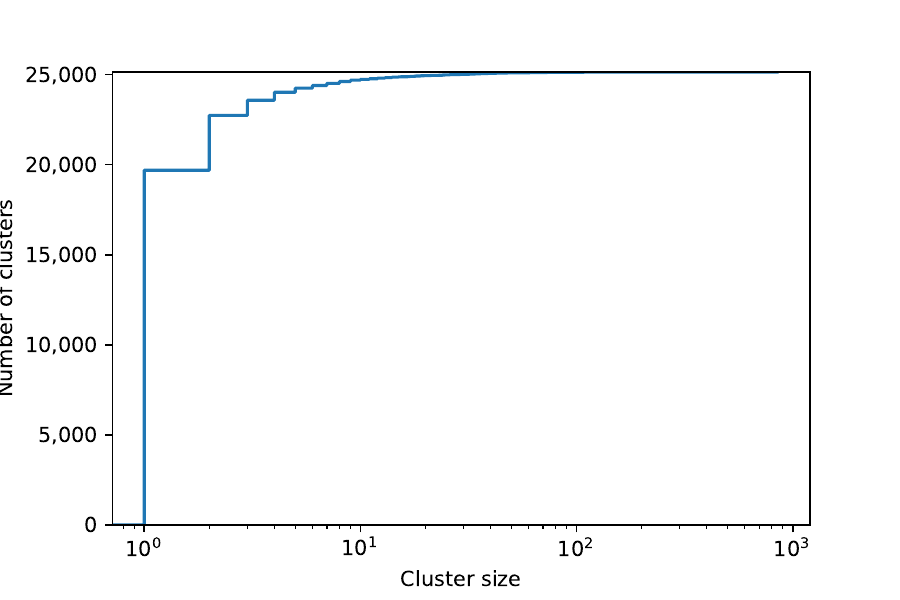}
    \caption{Left: The cumulative distribution function (CDF) comparing how many users (y-axis) submitted how many messages (x-axis) to the WhatsApp tiplines. While 54\% of users sent only one message, the most prolific user sent 299. Right: The CDF comparing the number of clusters (y-axis) to their sizes (x-axis) on the WhatsApp tiplines. While 78\% of clusters have only one item, the largest cluster has 858 instances of people thanking fact-checkers and the second largest has 210 instances of a video. Note that x-axes for both plots use log-10 scales. }
    \label{fig:tipline-userdist}
\end{figure}

The submissions were made by 14,959 unique users. 
The number of unique items submitted per tipline user appears to be heavy-tailed (Figure \ref{fig:tipline-userdist} left). While 54\% of users each submitted only one piece of content, the most active tipline user submitted 299 content items. Overall only 5\% of users are responsible for 41\% of the unique items submitted to the tiplines.

\begin{figure}[tb]
    \centering
    \includegraphics[width=0.5\textwidth]{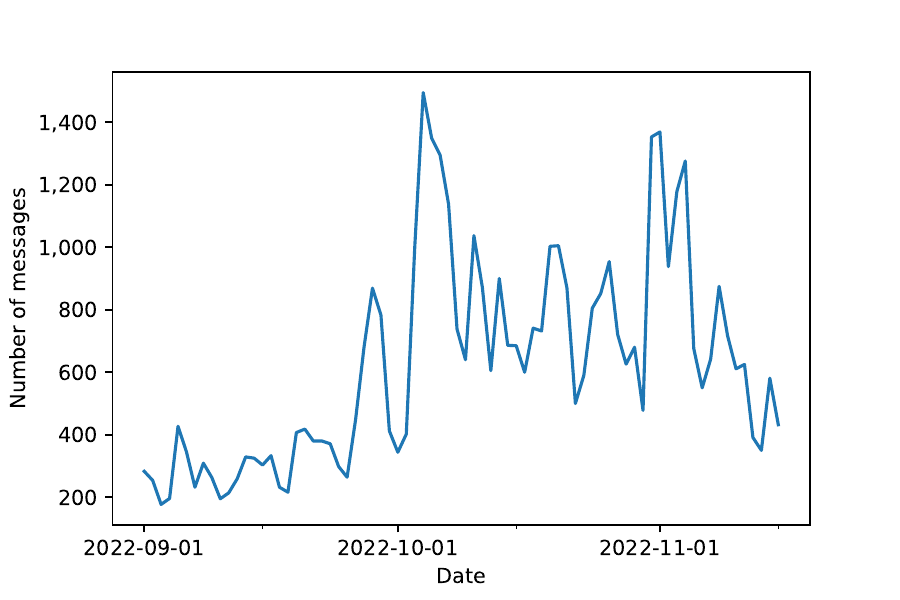}%
    \includegraphics[width=0.5\textwidth]{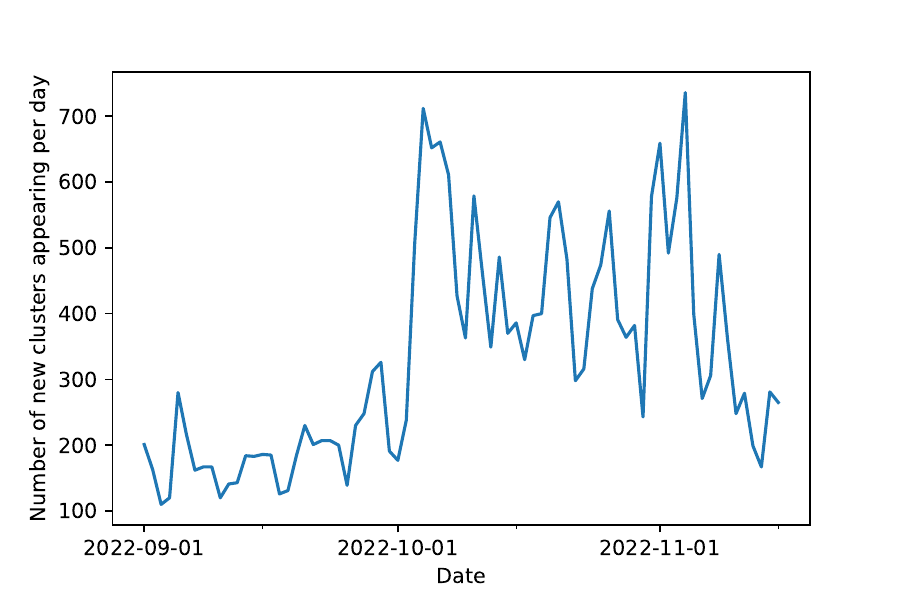}
    \caption{Left: Number of tipline submissions per day. Right: Number of new clusters appearing per day in the fact-checkers' misinformation tiplines.}
    \label{fig:number_of_submissions_per_day}
\end{figure}

We find the number of submissions per day to the tiplines varied considerably (Figure \ref{fig:number_of_submissions_per_day} left). There were pronounced peaks on the day of the two elections as well as another peak just after the run-off election. Manual inspection of content during this post-election peak revealed it was largely people inquiring about the results. The amount of novel content---represented by the number of new clusters (Figure \ref{fig:number_of_submissions_per_day} right)---follows a similar pattern to the number of daily messages, suggesting new messages often brought new content.

In general, we found little overlap between the users of different tiplines. Of the 6,383 users who submitted two or more messages, 93\% submitted all their messages to the same tipline, 7\% submitted at least one message to two tiplines, and 0.6\% submitted messages to all three tiplines.

\begin{figure}[tb]
    \centering
    \includegraphics[width=0.5\textwidth]{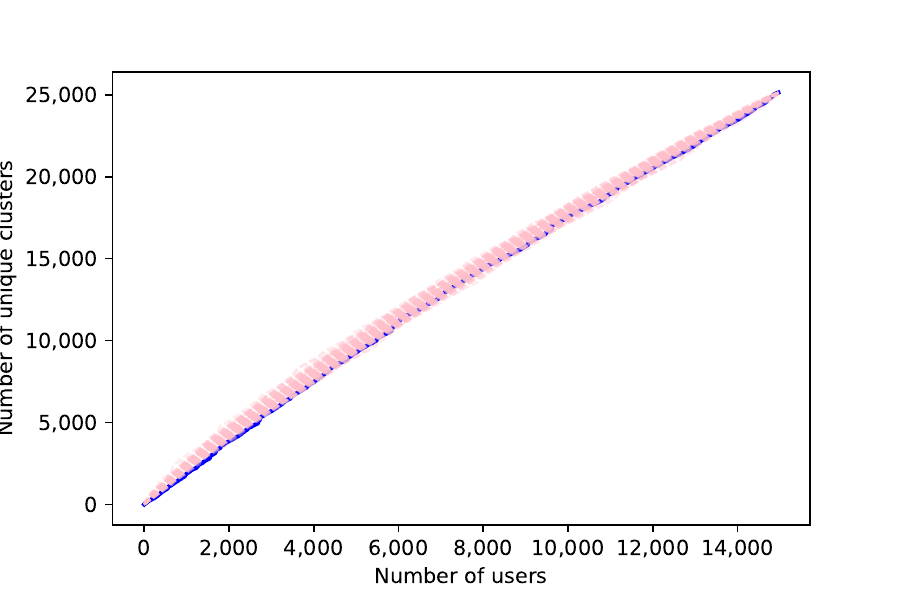}%
    \includegraphics[width=0.5\textwidth]{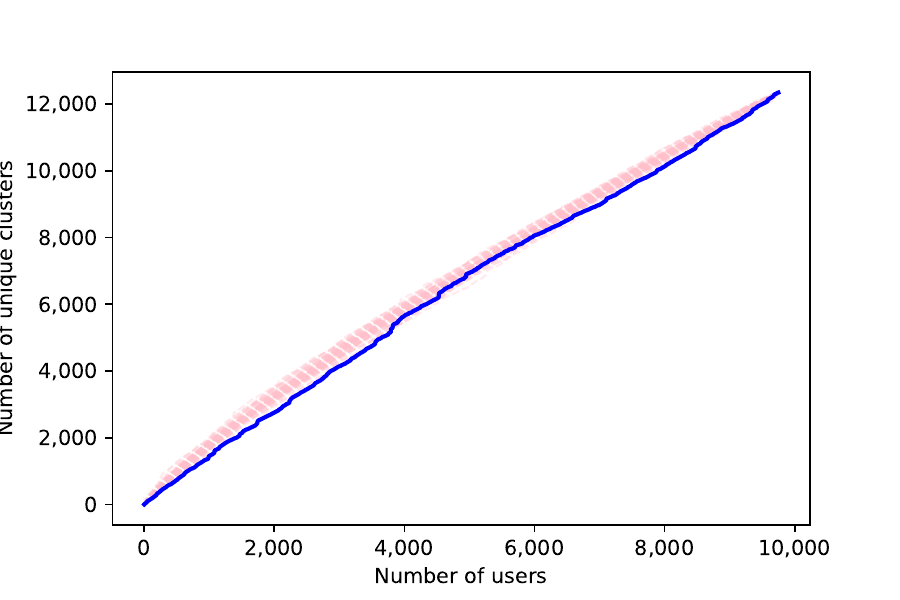}
    \caption{The relationship between the number of users and the amount of novel content (number of clusters) is mostly linear in the empirical data (solid blue line) and in random re-orders (pink dashed lines). The relationship is similar when consider all content (left) or only content that leads to a published fact-check (right).}
    \label{fig:shuffle}
\end{figure}

We find that novel textual content grows approximately linearly with the number of unique users (Figure~\ref{fig:shuffle}). The empirically observed distribution is shown in blue, and 100 random re-orderings are shown in dashed pink lines. This relationship holds when consider all items (left) or only items that were fact-checked (right). As discussed further in Section \ref{discussion}, the linear relationship suggests we are far from reaching a saturation point in content and more tipline users would likely lead to new content that would not otherwise be observed.

We manually examined the largest clusters of messages and found they were often video or long, forwarded text messages. We discarded a few large clusters that were greetings (e.g., `Bom dia'), expressions of gratitude to the fact-checkers, or spam. 
The largest remaining cluster is 210 copies of a video falsely alleging that one's vote will not be counted if they press ``Confirm vote'' too quickly on at the confirmation stage on the voting machine. 

The next largest cluster is of 175 similar text messages. The messages in this cluster are on average 3,945 characters long and tell a `first-hand account' of alleged corruption and bribing of the TSE without which Bolsonaro would have supposedly won the election.

There are further clusters of videos making false allegations about discarded votes in Curitiba (134 videos) or fraudulent ballots being prepared by a political party in advance (131 videos).
Then 122 requests are about video where an influencer and self-titled leader of the New Aeon Church of Lucifer claims that different religions and entities linked to Satanism and Luciferianism had come together to guarantee the PT's victory in the first round of voting.

The next largest cluster is one of 114 text messages alleging that ``dead people voted for Lula.'' The average length of the messages in this cluster was 807 characters. After the initial allegation, the messages listed the names of various cities, their number of inhabitants, and the number of votes cast for Lula. In each case the number of votes cast for Lula was higher than the number of the inhabitants listed. The messages concluded saying ``there are 192 more Brazilian cities in which the dead resurrected to vote for Lula.''
Other large clusters of \textit{text} messages shared election polling predictions (94 messages), long lists of the good things Bolsonaro did for Brazil (77 messages), and  `news' that protests had been successful in creating a new court that would oversee the Supreme Federal Court or STF (74 messages).

\subsubsection{TSE tipline}
\label{tse-analysis}
The data sent to the fact-checking feature of the TSE WhatsApp bot had far more text queries than the misinformation tiplines operated by fact-checkers. Between September 1, 2022 and November 15, 2022 inclusive the TSE fact-checking feature received 223,621 queries. The majority, 83\% are text messages, while video account for 7\%, images for 6\% and audios for 3\%.  This is explained in part by the fact that the feature only accepted text messages when it first launched. Multimedia messages were accepted from September 21, 2022.

The text messages sent to the TSE bot are also significantly shorter on average compared to those sent to the fact-checker tiplines. While the fact-checker tipline text messages have an average length of 637 characters (sd: 1152), the text messages sent to the TSE bot have an average of 35 characters (sd: 57). In contrast to the long messages sent to fact-checkers, the TSE messages often reassemble search queries. The largest cluster is short variations of `ballot security'  (`segurança das urnas') with 7,211 messages, but it must be noted that this was given as an example query by the bot, and some people clearly copied and pasted the example to try out the bot. In some instances the quotations marks around the phrase or the word `exemplo' [example] are included. Nevertheless, there are 204 variations of this query suggesting there was also genuine interest. There are also 956 searches for `fraude nas urnas' [ballot fraud].  
The other top clusters include `[e-]título' (``Voter ID'', 3,084 messages), `justificar voto' (``justifying vote'', 1,997 messages), `local de votação' (``place of voting'', 1,762), and `voto em trânsito' (``vote in transit'', 1,572). Despite being told the fact-checking feature was exclusively for concerns about the integrity of the elections, there are messages about candidates as well: `Lula' was submitted 857 times and `Bolsonaro' 780 times. 
Variations of `Lula ladrão' [Lula thief] and `Lula é inocente?' [Is Lula innocent] were sent 2,858 times. After the elections, searches for `resultado' [result] increased with 751 queries overall.

\begin{figure}
    \centering
    \includegraphics[width=0.7\textwidth]{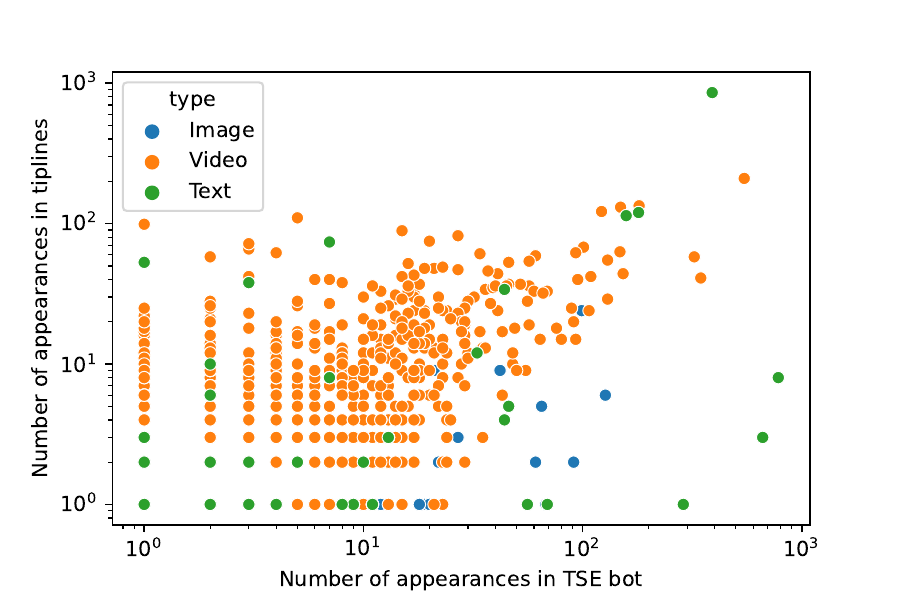}%
    \caption{Overlap between content submitted to the tiplines and TSE is low, but there is a weak, positive correlation between the number of times a video, image, or text item is sent to the TSE bot and the misinformation tiplines.}
    \label{fig:tipline_tse_scatter}
\end{figure}

Overall, overlap with the fact-checking misinformation tiplines is low. Of the 8,389 videos appearing across the tiplines and TSE bot, only 1,533 (18\%) appear in both. Similarly only 1\% of images and less than 0.01\% of text claims appear in both sources.
For the items appearing in both the tiplines and the TSE bot, there is a weak, but positive and statistically significant correlation between the number of appearances in both (Figure \ref{fig:tipline_tse_scatter}). The correlation is 0.67 for videos, 0.55 for images and 0.32 for text messages.

\begin{figure}[bt]
    \centering
    \includegraphics[width=0.5\textwidth]{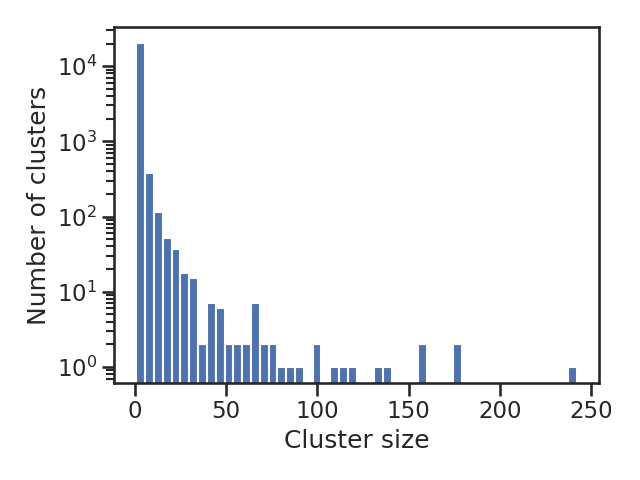}%
    \includegraphics[width=0.6\textwidth]{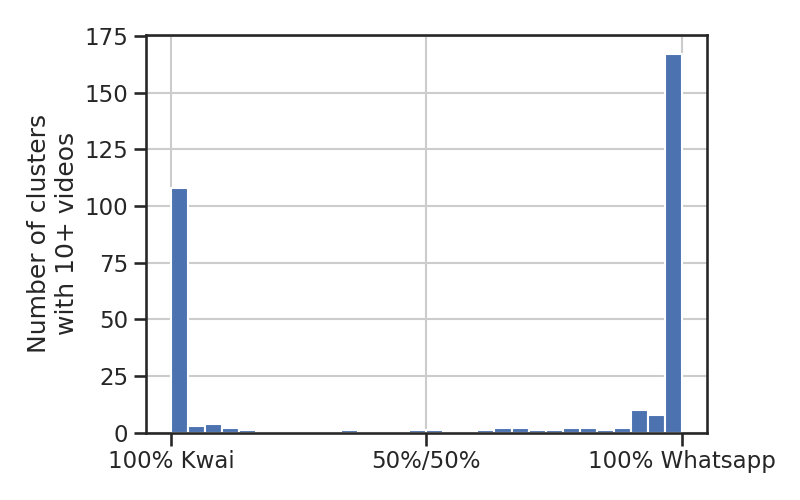}
    \caption{\textbf{(left)} Distribution of cluster sizes for the clusters of TMK embeddings of videos from WhatsApp and Kwai. \textbf{(right)} Histogram of the percentage of WhatsApp videos per cluster, for the 320 clusters with 10 or more videos.}
    \label{fig:whatsapp-vs-kwai}
\end{figure}

\subsection{Kwai and WhatsApp}
\label{sec:kwai_data_analysis}

We identified 13,068 unique Kwai users. As in WhatsApp data, the distribution of posts per user is heavy-tailed: 82\% of the users appear in our dataset with just one post and 10\% with two. One user had with 223 videos in our dataset and ran a channel exclusively dedicated to the Brazilian elections.

In our data collection, we find that Kwai suspended the search results for some terms related to the Brazilian elections and confirmed findings reported by the Brazilian press.

On 12 and 24 October, respectively, the terms `urnas' [ballot] and `eleições' [elections] ceased to return results. Kwai users based in Brazil confirmed on 23 October that the Kwai mobile application also did not present search results for those terms. The changes happened after 9 October, when the Brazilian press reported that Kwai was suggesting misinformation and misleading content about the Brazilian elections \citep{galf_kwai_2022}.\footnote{According to \citet{galf_kwai_2022}, the hashtag \#fraudenasurnas [ballot fraud] reached at least 1.4 million views and the auto-complete search feature in Kwai offered suggestions like ``urnas eletrônicas fraudadas" [rigged electronic ballot boxes] for the term ``urna" [ballot] or ``fraude nas urnas eletrônicas 2022" [electronic ballot box fraud 2022] for the term `fraude' [fraud]. \citet{scofield_tiktok_2022} also noted the existence of videos containing misinformation about the elections in Kwai.} We gathered 197 videos referenced as search results for `urnas' [ballots] between 8 and 11 October 2022. Analysis of the top 20 videos ranked by view count, conducted by our team, revealed that 15 of them contain misleading information or were created to spread doubt about the integrity of the electoral process.

Proceeding to the text analysis, we embed the video descriptions and manually inspect the largest clusters. We find 9,068 semantic clusters for the 15,017 video descriptions. The largest cluster contains 301 posts, and its keywords mirror the broad topic of our research (``Eleicoes2022" and variations). The second most important cluster identified has 85 posts and contains words associated with the Brazilian army (``exercito"). For comparison purposes, the top clusters associated with the words ``Lula" and ``Bolsonaro" had 55 and 54 posts, respectively.

To compare the video content present on the WhatsApp and Kwai datasets, we produced TMK embeddings for all 35,701 videos from both data sources, and used the \texttt{tmk-clusterize} executable to cluster those embeddings, with the recommended 0.7 threshold for both level 1 and level 2 similarity measures. We find that increasing the threshold to a more stringent value of 0.9 does not result in any relevant difference in our results suggesting that the findings are robust to a range of threshold choices.
In total, we find 20,887 clusters, of which 17,364 are singleton clusters, i.e. clusters with only one video. Of those 17,364 singleton clusters, 6,338 videos are from WhatsApp, and 11,026 videos are from Kwai (a 36\%/64\% split). The left panel of Figure~\ref{fig:whatsapp-vs-kwai} shows a histogram of cluster size for all clusters. 

For the non-singleton clusters, we find that the majority of clusters are composed of either nearly 100\% of Kwai videos, or of nearly 100\% of WhatsApp videos.
This is illustrated on the right panel of Figure~\ref{fig:whatsapp-vs-kwai}, which shows a histogram of the percentage of WhatsApp videos per cluster for the 320 clusters with 10 or more videos: the distribution is bimodal, with the two largest peaks corresponding to clusters that are composed of videos exclusive to Kwai (corresponding to 32.2\% of clusters)  or exclusive to WhatsApp (49.7\% of clusters). 

\subsection{Twitter and WhatsApp}
We find embeddings and existing claim matching algorithms have low precision for claim matching the long WhatsApp text messages with short Twitter posts. As a result, we use a combination of quantitative methods and manual analysis to investigate the prevalence of the top four WhatsApp text messages on Twitter. The top cluster of text messages on WhatsApp alleges corruption at the Ministry of Justice or the TSE. The messages in this cluster made use of phrases such as, `se as pessoas soubessem o que aconteceu nos bastidores do TSE ficariam enojadas' [If people knew what happened behind the scenes at TSE they would be disgusted] or, `se as pessoas soubessem o que aconteceu no ministerio da justica ficariam enojadas' [If people knew what happened at the Ministry of Justice they would be disgusted.' 

The expression ``If people knew X, they would be disgusted'' is a long-standing \textit{snowclone} (a phrasal template or, effectively, a meme) in Brazil.\footnote{The term ``snowclone'' was introduced by American linguists Geoffrey K. Pullum and Glen Whitman to describe textual templates and cliché frames~\citep{upennLanguageLog}.} It dates back to misinformation claims related to the 1998 World Cup and has been adapted by Brazilian Internet users in various contexts since then. Most tweets in this cluster were short, often repeating the sentence with little variation and no further detail---probably due to the meme status of the sentence. We examined the results returned by the semantic models and then filtered them to only those containing `enojadas' and `pessoas soubessem.' With this filtering, we found 23 matching tweets. 
The oldest matching tweet in our data was published on Oct 3, 2022, while the first occurrence of the message in the WhatsApp tiplines was on September 5. While the matching tweets often did not provide further detail, the WhatsApp messages are longer and contain lists of claims about bribes, corruption, and other unsubstantiated assertions.

\begin{figure}
    \centering
    \includegraphics[width=0.5\textwidth]{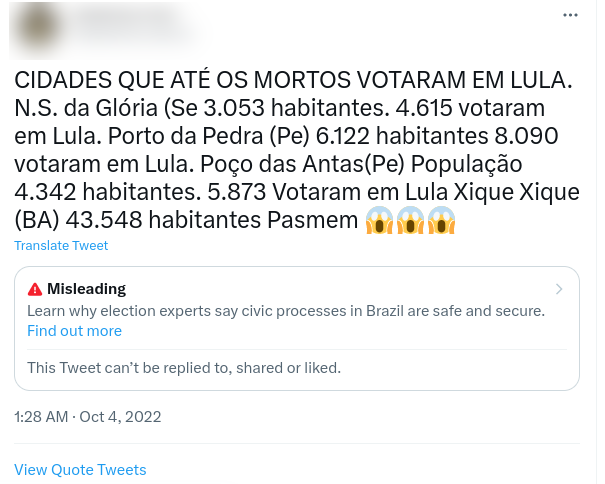}%
    \includegraphics[width=0.5\textwidth]{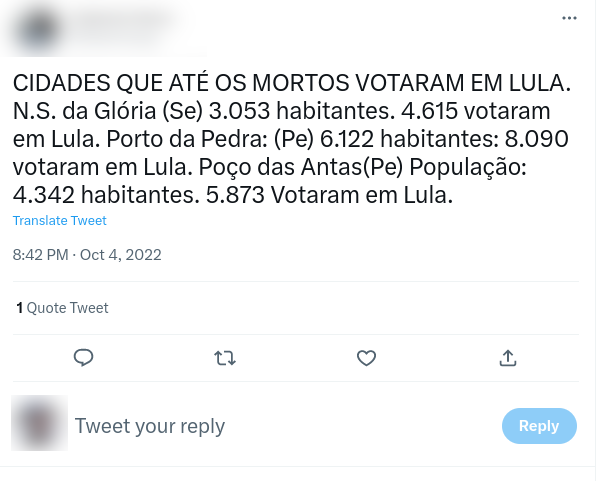}
    \caption{Tweets alleging there where ``cities where even the dead voted for Lula.'' While some tweets were labeled as misleading (left), others were not (right). }
    \label{fig:labeled_tweet}
\end{figure}

The second largest cluster of WhatsApp text messages is about `cidades que ate os mortos votaram em lula' [cities where even the dead voted for Lula]. After manual inspection, we filtered messages returned with the semantic models to only those containing `mortos' [dead] and `votar' [vote] and found 284 matching tweets. The earliest tweet was published at 1:28 AM UTC on Oct 4, 2022, while the earliest submission on the WhatsApp tipline was sent a little less than an hour earlier at 00:48 AM UTC on the same day. While the first tweet on Twitter was labeled as `misleading' and could no longer be found when searching for `cidades que ate os mortos votaram em lula' on the Twitter website despite that exact phrase appearing in the tweet. Nonetheless, at the time of writing in January 2023 there were many very similar tweets with the same phrase that were not labeled as misleading (Figure \ref{fig:labeled_tweet}). One tweet accessible via the search feature on Twitter includes screenshots of the messages on WhatsApp.

The third largest WhatsApp cluster was a message describing (incorrect) partial poll results in each Brazilian state, and asking other WhatsApp users to share the content with their friends and Bolsonaro supporters. It claims that mainstream news sources and social media platforms would not publicize the results otherwise. The matches in this cluster are mixed: the top matches in the cluster, i.e. the tweets with the highest similarity to the WhatsApp message, state support for Bolsonaro but do not mention any polls. Further matches, on the other hand, mention poll results but do not always state support for a candidate. This divide is a good illustration of the challenge in clustering embeddings of messages of variable length: since many of the longer WhatsApp messages contain a header or a call to action (e.g. ``everybody please read this'', ``please share this message''), as well as the other contextual elements such as expressing support for a candidate, semantic similarity-induced clusters are likely to reflect these characteristics of the messages, rather than only producing topic-centered clusters.

Finally, the fourth largest cluster is centered around a WhatsApp message listing Bolsonaro's successes as president (with varying degrees of accuracy throughout the list). The full WhatsApp message lists 60 accomplishments and has a length of over 3,600 characters, which is more than ten times the limit of 280 character limit for a tweet. In this case, even though no single tweet accurately matches the whole list from the WhatsApp messages, a Twitter search for parts of the list returns multiple tweets containing the individual claims on the list, as well as Twitter threads containing multiple tweets, each with one or a few claims from the list. This cluster reveals another challenge of matching different social media platforms such as Twitter and WhatsApp: trying to match long messages from one platform with short posts on another. While we focused on individual messages as our unit of analysis, at times one WhatsApp message can be more like an entire Twitter thread of self-replies.

\section{Discussion}
\label{discussion}
The central focus of our investigation was to understand how misinformation manifests and circulates across different social media platforms during the 2022 Brazilian general election. Affordance theory \citep{gibson2014theory} argues that the specific features and perceived utilities of a platform will influence its user interactions and the nature of content that circulates. In contrast, the fact that people use multiple platforms in everyday life provides a way for similar content to spread across platforms. 
Applying this to our observations, it appears that the unique characteristics of Twitter, WhatsApp, and Kwai provide an ecosystem for specific types or formats of misinformation but that there is some degree of overlap. %

While much scholarship focuses on Twitter and Western countries, this paper attempts to describe misinformation on three platforms during the 2022 Brazilian general election and answer research questions about the overlap. We find that there are examples of common misinformation claims across WhatsApp, Kwai, and Twitter, but that the claims are presented differently on each platform. This appears to be due, at least in part, to the unique characteristics of each platform as predicted by affordance theory \citep{gibson2014theory}. For instance, the top WhatsApp text messages submitted to the misinformation tiplines operated by fact-checkers are often order of magnitude larger than the maximum size allowed for tweets.

These differences also create challenges for claim matching algorithms \citep{kazemi_claim_2021,shaar_that_2020} to match similar claims across different platforms and complicate large-scale quantitative analysis. The precision for claim matching with vector models alone was so low that we needed to employ additional keyword filtering and manual analysis to understand the overlap between Twitter and WhatsApp text messages. As a result, we limited our analysis to examining the prevalence of the top WhatsApp misinformation claims on Twitter. 
Further research is needed to develop claim matching algorithms that work across platforms, are robust to stylistic differences (e.g., the use of accents, diacritics, and capitalization), and handle common abbreviations used on social media.

The most complete, large-scale comparison we can conduct is between the videos on Kwai and WhatsApp.
There is limited overlap between the videos on each platform, and most content appears on only one of the two platforms, with 32.2\% of clusters with 10 videos or more containing videos only found on Kwai, and 49.7\% containing videos only found on WhatsApp.

This difference between content found on Kwai vs.\ WhatsApp might come as a surprise to WhatsApp researchers interested in Brazilian misinformation networks who annecdotally observe Kwai videos circulating on WhatsApp; however, it is important to bear in mind that the WhatsApp dataset we analyze is not a general sample of WhatsApp. Our data is specifically a crowdsourced sample of content sent to fact-checking tiplines. %
Additionally, since Kwai does not have an official API, we could not access all posts in the platform, only those returned as search results for our query terms. Therefore, it is possible that Kwai videos shared in the tiplines have not been matched because they were not returned as search results for our queries.
Our study cannot explain the mechanisms that lead to different content on Kwai and WhatsApp, but the theory in Section \ref{sec:literature} suggests the affordances users perceive on each platform and the audiences they imagine consuming their content may play a role. %

Our comparative analysis of WhatsApp tiplines revealed that most of the users submit only one message, although a smaller number of `power-users' submit many messages. This mirrors the heavy-tailed patterns found for content creation on other platforms \citep[e.g.,][]{panciera2009wikipedians}. Of the users who submit multiple messages, most interact exclusively with one tipline, suggesting each fact-checking organization serves a different audience. Given the end-to-end encryption of WhatsApp, it's impossible to know how representative any collection of messages is. Our analysis suggests that new claims are constantly circulating: the amount of unique content is highly correlated with the number of users.
It thus appears that the tiplines are not reaching a saturation point: as the number of users interacting with a tipline increases so too does the amount of novel, unseen content.

\section{Conclusion} \label{sec:conclusion}

The cross-platform movement of content, while evident in our data, does not lead to a homogeneous set of misinformation claims across platforms. Even when content does migrate from one platform to another, it is often reshaped or recontextualized to fit the new platform's affordances. For example, messages may change length or format and be tailored to different audiences. %
These findings have important practical implications for fact-checkers and academics. Fact-checking organizations need to adopt approaches to discovering misinformation claims relevant to their audiences on multiple platforms. It is not sufficient to monitor only the `easy-to-access' platforms and focus on the misinformation claims there. Especially in Brazil where WhatsApp use is far greater than Twitter, it is necessary to develop approaches to identify misinformation claims directly on the platform. Thankfully several options are developing: Fact-checking organizations can run tiplines like those used in this paper, scrape content from large, `public' groups \citep{garimella2018whatsapp}, and use innovative data donation approaches such as the WhatsViral Android App that hashes content on device.\footnote{https://play.google.com/store/apps/details?id=com.rutgers.whatsviral}

For researchers, the results we present here show promise, but also highlight many opportunities for better algorithms and tools to aggregate multimodal content at scale and in meaningful ways. The growth of video-based platforms such as Kwai and TikTok, as well as of video forms of content such as Instagram Reels, demands a different set of analytical tools than those for text-based platforms. While TMK embeddings have proven to be useful in the identification of the same videos across platforms, they do not address semantic elements of video content, i.e.\ when videos are on the same topic, or frame an issue in the same way. Moreover, TMK embeddings use the visual part of the videos, whereas one could argue for a clustering based on whether videos share the same audio track, or perhaps the same voice or speaker. 

Our results also show the promise of crowdsourcing approaches to identifying misinformation on end-to-end encrypted messaging platforms. Partnering with fact-checking organizations to better understand the content circulating on such platforms and to develop tools to assist fact-checkers to respond is an important area for further research.
In addition, our results highlight the potential of Kwai as a data source for further research. It provides a valuable window into the online content consumed by low-income users in Brazil, who are traditionally underrepresented in online data studies based on platforms such as Twitter. %

Overall, we hope that our research encourages further analysis of new social media platforms, particularly those used in the Global South, while also encouraging more multiplatform studies and studies in co-operation with fact-checking organizations. To support this, we need to develop algorithms that can extract and match claims and perform other tasks across different platforms.


\begin{thebibliography}{}

\bibitem [\protect \citeauthoryear {%
Altay%
, Berriche%
\BCBL {}\ \BBA {} Acerbi%
}{%
Altay%
\ \protect \BOthers {.}}{%
{\protect \APACyear {2023}}%
}]{%
altay2023misinformation}
\APACinsertmetastar {%
altay2023misinformation}%
\begin{APACrefauthors}%
Altay, S.%
, Berriche, M.%
\BCBL {}\ \BBA {} Acerbi, A.%
\end{APACrefauthors}%
\unskip\
\newblock
\APACrefYearMonthDay{2023}{}{}.
\newblock
{\BBOQ}\APACrefatitle {Misinformation on Misinformation: Conceptual and
  Methodological Challenges} {Misinformation on misinformation: Conceptual and
  methodological challenges}.{\BBCQ}
\newblock
\APACjournalVolNumPages{Social Media + Society}{9}{1}{20563051221150412}.
\newblock
\begin{APACrefURL} \url{https://doi.org/10.1177/20563051221150412}
  \end{APACrefURL}
\newblock
\begin{APACrefDOI} \doi{10.1177/20563051221150412} \end{APACrefDOI}
\PrintBackRefs{\CurrentBib}

\bibitem [\protect \citeauthoryear {%
Anita~Baptista%
, Rossini%
, Veiga~de Oliveira%
\BCBL {}\ \BBA {} Stromer-Galley%
}{%
Anita~Baptista%
\ \protect \BOthers {.}}{%
{\protect \APACyear {2019}}%
}]{%
anita2019circulaccao}
\APACinsertmetastar {%
anita2019circulaccao}%
\begin{APACrefauthors}%
Anita~Baptista, E.%
, Rossini, P.%
, Veiga~de Oliveira, V.%
\BCBL {}\ \BBA {} Stromer-Galley, J.%
\end{APACrefauthors}%
\unskip\
\newblock
\APACrefYearMonthDay{2019}{}{}.
\newblock
{\BBOQ}\APACrefatitle {A circula{\c{c}}{\~a}o da (des) informa{\c{c}}{\~a}o
  pol{\'\i}tica no {WhatsApp} e no {Facebook}} {A circula{\c{c}}{\~a}o da (des)
  informa{\c{c}}{\~a}o pol{\'\i}tica no {WhatsApp} e no {Facebook}}.{\BBCQ}
\newblock
\APACjournalVolNumPages{Lumina}{13}{3}{29--46}.
\PrintBackRefs{\CurrentBib}

\bibitem [\protect \citeauthoryear {%
{Asian American Disinformation Table}%
}{%
{Asian American Disinformation Table}%
}{%
{\protect \APACyear {2022}}%
}]{%
disinfotable2022power}
\APACinsertmetastar {%
disinfotable2022power}%
\begin{APACrefauthors}%
{Asian American Disinformation Table}.%
\end{APACrefauthors}%
\unskip\
\newblock
\APACrefYearMonthDay{2022}{}{}.
\newblock
\APACrefbtitle {Power, Platforms, Politics: {A}sian {A}mericans and
  Disinformation Landscape Report.} {Power, platforms, politics: {A}sian
  {A}mericans and disinformation landscape report.}
\newblock
\begin{APACrefURL}
  \url{https://www.asianamdisinfo.org/wp-content/uploads/2022/08/AsianAmDisinformation_LandscapeReport2022.pdf}
  \end{APACrefURL}
\PrintBackRefs{\CurrentBib}

\bibitem [\protect \citeauthoryear {%
Banaji%
, Bhat%
, Agarwal%
, Passanha%
\BCBL {}\ \BBA {} Sadhana~Pravin%
}{%
Banaji%
\ \protect \BOthers {.}}{%
{\protect \APACyear {2019}}%
}]{%
banaji2019whatsapp}
\APACinsertmetastar {%
banaji2019whatsapp}%
\begin{APACrefauthors}%
Banaji, S.%
, Bhat, R.%
, Agarwal, A.%
, Passanha, N.%
\BCBL {}\ \BBA {} Sadhana~Pravin, M.%
\end{APACrefauthors}%
\unskip\
\newblock
\APACrefYearMonthDay{2019}{}{}.
\newblock
\APACrefbtitle {Whats{A}pp vigilantes: An exploration of citizen reception and
  circulation of {WhatsApp} misinformation linked to mob violence in India.}
  {Whats{A}pp vigilantes: An exploration of citizen reception and circulation
  of {WhatsApp} misinformation linked to mob violence in india.}
\newblock
\APACaddressPublisher{}{Department of Media and Communications, London School
  of Economics and Political Science}.
\newblock
\begin{APACrefURL}
  \url{https://blogs.lse.ac.uk/medialse/2019/11/11/whatsapp-vigilantes-an-exploration-of-citizen-reception-and-circulation-of-whatsapp-misinformation-linked-to-mob-violence-in-india/}
  \end{APACrefURL}
\PrintBackRefs{\CurrentBib}

\bibitem [\protect \citeauthoryear {%
Batista~Pereira%
, Bueno%
, Nunes%
\BCBL {}\ \BBA {} Pavão%
}{%
Batista~Pereira%
\ \protect \BOthers {.}}{%
{\protect \APACyear {2022}}%
}]{%
batista_pereira_fake_2022}
\APACinsertmetastar {%
batista_pereira_fake_2022}%
\begin{APACrefauthors}%
Batista~Pereira, F.%
, Bueno, N\BPBI S.%
, Nunes, F.%
\BCBL {}\ \BBA {} Pavão, N.%
\end{APACrefauthors}%
\unskip\
\newblock
\APACrefYearMonthDay{2022}{{\APACmonth{10}}}{}.
\newblock
{\BBOQ}\APACrefatitle {Fake {News}, {Fact} {Checking}, and {Partisanship}:
  {The} {Resilience} of {Rumors} in the 2018 {Brazilian} {Elections}} {Fake
  {News}, {Fact} {Checking}, and {Partisanship}: {The} {Resilience} of {Rumors}
  in the 2018 {Brazilian} {Elections}}.{\BBCQ}
\newblock
\APACjournalVolNumPages{The Journal of Politics}{84}{4}{}.
\newblock
\begin{APACrefURL}
  \url{https://www.journals.uchicago.edu/doi/full/10.1086/719419}
  \end{APACrefURL}
\newblock
\begin{APACrefDOI} \doi{10.1086/719419} \end{APACrefDOI}
\PrintBackRefs{\CurrentBib}

\bibitem [\protect \citeauthoryear {%
Belli%
}{%
Belli%
}{%
{\protect \APACyear {2017}}%
}]{%
belli_net_2017}
\APACinsertmetastar {%
belli_net_2017}%
\begin{APACrefauthors}%
Belli, L.%
\end{APACrefauthors}%
\unskip\
\newblock
\APACrefYearMonthDay{2017}{{\APACmonth{01}}}{}.
\newblock
{\BBOQ}\APACrefatitle {Net neutrality, zero rating and the {Minitelisation} of
  the internet} {Net neutrality, zero rating and the {Minitelisation} of the
  internet}.{\BBCQ}
\newblock
\APACjournalVolNumPages{Journal of Cyber Policy}{2}{1}{}.
\newblock
\begin{APACrefURL} \url{https://doi.org/10.1080/23738871.2016.1238954}
  \end{APACrefURL}
\newblock
\begin{APACrefDOI} \doi{10.1080/23738871.2016.1238954} \end{APACrefDOI}
\PrintBackRefs{\CurrentBib}

\bibitem [\protect \citeauthoryear {%
Belli%
\ \protect \BOthers {.}}{%
Belli%
\ \protect \BOthers {.}}{%
{\protect \APACyear {2018}}%
}]{%
belli_governance_2018}
\APACinsertmetastar {%
belli_governance_2018}%
\begin{APACrefauthors}%
Belli, L.%
, Cavalli, O.%
, Álvarez, C.%
, Arellano, P\BPBI B.%
, Azzolin, H.%
, Baca-Feldman, C\BPBI F.%
\BDBL {}Zingales, N.%
\end{APACrefauthors}%
\unskip\
\newblock
\APACrefYear{2018}.
\newblock
\APACrefbtitle {Governança e regulações da {Internet} na {América}
  {Latina}: análise sobre infraestrutura, privacidade, cibersegurança e
  evoluções tecnológicas em homenagem aos dez anos da {South} {School} on
  {Internet} {Governance}} {Governança e regulações da {Internet} na
  {América} {Latina}: análise sobre infraestrutura, privacidade,
  cibersegurança e evoluções tecnológicas em homenagem aos dez anos da
  {South} {School} on {Internet} {Governance}}.
\newblock
\begin{APACrefURL}
  \url{http://bibliotecadigital.fgv.br/dspace/handle/10438/27164}
  \end{APACrefURL}
\PrintBackRefs{\CurrentBib}

\bibitem [\protect \citeauthoryear {%
Blank%
, Dutton%
\BCBL {}\ \BBA {} Lefkowitz%
}{%
Blank%
\ \protect \BOthers {.}}{%
{\protect \APACyear {2020}}%
}]{%
blank2020oxis}
\APACinsertmetastar {%
blank2020oxis}%
\begin{APACrefauthors}%
Blank, G.%
, Dutton, W\BPBI H.%
\BCBL {}\ \BBA {} Lefkowitz, J.%
\end{APACrefauthors}%
\unskip\
\newblock
\APACrefYearMonthDay{2020}{}{}.
\newblock
{\BBOQ}\APACrefatitle {OxIS 2019: The rise of mobile internet use in Britain}
  {Oxis 2019: The rise of mobile internet use in britain}.{\BBCQ}
\newblock
\APACjournalVolNumPages{Available at SSRN 3538301}{}{}{}.
\PrintBackRefs{\CurrentBib}

\bibitem [\protect \citeauthoryear {%
Brummette%
, DiStaso%
, Vafeiadis%
\BCBL {}\ \BBA {} Messner%
}{%
Brummette%
\ \protect \BOthers {.}}{%
{\protect \APACyear {2018}}%
}]{%
brummette2018politicization}
\APACinsertmetastar {%
brummette2018politicization}%
\begin{APACrefauthors}%
Brummette, J.%
, DiStaso, M.%
, Vafeiadis, M.%
\BCBL {}\ \BBA {} Messner, M.%
\end{APACrefauthors}%
\unskip\
\newblock
\APACrefYearMonthDay{2018}{}{}.
\newblock
{\BBOQ}\APACrefatitle {Read All About It: The Politicization of “Fake News”
  on Twitter} {Read all about it: The politicization of “fake news” on
  twitter}.{\BBCQ}
\newblock
\APACjournalVolNumPages{Journalism \& Mass Communication
  Quarterly}{95}{2}{497-517}.
\newblock
\begin{APACrefURL} \url{https://doi.org/10.1177/1077699018769906}
  \end{APACrefURL}
\newblock
\begin{APACrefDOI} \doi{10.1177/1077699018769906} \end{APACrefDOI}
\PrintBackRefs{\CurrentBib}

\bibitem [\protect \citeauthoryear {%
Cetic.br%
}{%
Cetic.br%
}{%
{\protect \APACyear {2021}}%
}]{%
ceticbr_survey_2021}
\APACinsertmetastar {%
ceticbr_survey_2021}%
\begin{APACrefauthors}%
Cetic.br.%
\end{APACrefauthors}%
\unskip\
\newblock
\APACrefYearMonthDay{2021}{}{}.
\newblock
\APACrefbtitle {Survey on the {Use} of {Information} and {Communication}
  {Technologies} in {Brazilian} {Households} - {ICT} {Households} 2021} {Survey
  on the {Use} of {Information} and {Communication} {Technologies} in
  {Brazilian} {Households} - {ICT} {Households} 2021}\ \APACbVolEdTR{}{\BTR{}}.
\newblock
\begin{APACrefURL}
  \url{https://cetic.br/publicacao/executive-summary-survey-on-the-use-of-information-and-communication-technologies-in-brazilian-households-ict-households-2021}
  \end{APACrefURL}
\PrintBackRefs{\CurrentBib}

\bibitem [\protect \citeauthoryear {%
Chauchard%
\ \BBA {} Garimella%
}{%
Chauchard%
\ \BBA {} Garimella%
}{%
{\protect \APACyear {2022}}%
}]{%
chauchard2022what}
\APACinsertmetastar {%
chauchard2022what}%
\begin{APACrefauthors}%
Chauchard, S.%
\BCBT {}\ \BBA {} Garimella, K.%
\end{APACrefauthors}%
\unskip\
\newblock
\APACrefYearMonthDay{2022}{Mar.}{}.
\newblock
{\BBOQ}\APACrefatitle {What Circulates on Partisan {WhatsApp} in {India}?
  Insights from an Unusual Dataset} {What circulates on partisan {WhatsApp} in
  {India}? insights from an unusual dataset}.{\BBCQ}
\newblock
\APACjournalVolNumPages{Journal of Quantitative Description: Digital
  Media}{2}{}{}.
\newblock
\begin{APACrefURL} \url{https://journalqd.org/article/view/2690}
  \end{APACrefURL}
\newblock
\begin{APACrefDOI} \doi{10.51685/jqd.2022.006} \end{APACrefDOI}
\PrintBackRefs{\CurrentBib}

\bibitem [\protect \citeauthoryear {%
Cihon%
\ \BBA {} Yasseri%
}{%
Cihon%
\ \BBA {} Yasseri%
}{%
{\protect \APACyear {2016}}%
}]{%
cihon2016twitter}
\APACinsertmetastar {%
cihon2016twitter}%
\begin{APACrefauthors}%
Cihon, P.%
\BCBT {}\ \BBA {} Yasseri, T.%
\end{APACrefauthors}%
\unskip\
\newblock
\APACrefYearMonthDay{2016}{}{}.
\newblock
{\BBOQ}\APACrefatitle {A Biased Review of Biases in Twitter Studies on
  Political Collective Action} {A biased review of biases in twitter studies on
  political collective action}.{\BBCQ}
\newblock
\APACjournalVolNumPages{Frontiers in Physics}{4}{}{}.
\newblock
\begin{APACrefURL}
  \url{https://www.frontiersin.org/articles/10.3389/fphy.2016.00034}
  \end{APACrefURL}
\newblock
\begin{APACrefDOI} \doi{10.3389/fphy.2016.00034} \end{APACrefDOI}
\PrintBackRefs{\CurrentBib}

\bibitem [\protect \citeauthoryear {%
Deck%
\ \BBA {} Marasciulo%
}{%
Deck%
\ \BBA {} Marasciulo%
}{%
{\protect \APACyear {2022}}%
}]{%
deck_tiktoks_2022}
\APACinsertmetastar {%
deck_tiktoks_2022}%
\begin{APACrefauthors}%
Deck, A.%
\BCBT {}\ \BBA {} Marasciulo, M.%
\end{APACrefauthors}%
\unskip\
\newblock
\APACrefYearMonthDay{2022}{{\APACmonth{03}}}{}.
\newblock
{\BBOQ}\APACrefatitle {{TikTok}’s biggest {Chinese} competitor bets big on
  {Brazil}} {{TikTok}’s biggest {Chinese} competitor bets big on
  {Brazil}}.{\BBCQ}
\newblock
\APACjournalVolNumPages{Rest of World}{}{}{}.
\newblock
\begin{APACrefURL}
  \url{https://restofworld.org/2022/tiktok-competitor-kwai-brazil/}
  \end{APACrefURL}
\PrintBackRefs{\CurrentBib}

\bibitem [\protect \citeauthoryear {%
Dwoskin%
}{%
Dwoskin%
}{%
{\protect \APACyear {2023}}%
}]{%
dwoskin_come_2023}
\APACinsertmetastar {%
dwoskin_come_2023}%
\begin{APACrefauthors}%
Dwoskin, E.%
\end{APACrefauthors}%
\unskip\
\newblock
\APACrefYearMonthDay{2023}{{\APACmonth{01}}}{}.
\newblock
{\BBOQ}\APACrefatitle {Come to the ‘war cry party’: {How} social media
  helped drive mayhem in {Brazil}} {Come to the ‘war cry party’: {How}
  social media helped drive mayhem in {Brazil}}.{\BBCQ}
\newblock
\APACjournalVolNumPages{Washington Post}{}{}{}.
\newblock
\begin{APACrefURL}
  \url{https://www.washingtonpost.com/technology/2023/01/08/brazil-bolsanaro-twitter-facebook/}
  \end{APACrefURL}
\PrintBackRefs{\CurrentBib}

\bibitem [\protect \citeauthoryear {%
Evangelista%
\ \BBA {} Bruno%
}{%
Evangelista%
\ \BBA {} Bruno%
}{%
{\protect \APACyear {2019}}%
}]{%
evangelista_whatsapp_2019}
\APACinsertmetastar {%
evangelista_whatsapp_2019}%
\begin{APACrefauthors}%
Evangelista, R.%
\BCBT {}\ \BBA {} Bruno, F.%
\end{APACrefauthors}%
\unskip\
\newblock
\APACrefYearMonthDay{2019}{}{}.
\newblock
{\BBOQ}\APACrefatitle {{WhatsApp} and political instability in {Brazil}:
  targeted messages and political radicalisation} {{WhatsApp} and political
  instability in {Brazil}: targeted messages and political
  radicalisation}.{\BBCQ}
\newblock
\APACjournalVolNumPages{Internet Policy Review}{8}{4}{1--23}.
\newblock
\begin{APACrefURL} \url{https://www.econstor.eu/handle/10419/214094}
  \end{APACrefURL}
\newblock
\begin{APACrefDOI} \doi{10.14763/2019.4.1434} \end{APACrefDOI}
\PrintBackRefs{\CurrentBib}

\bibitem [\protect \citeauthoryear {%
Galf%
}{%
Galf%
}{%
{\protect \APACyear {2022}}%
}]{%
galf_kwai_2022}
\APACinsertmetastar {%
galf_kwai_2022}%
\begin{APACrefauthors}%
Galf, R.%
\end{APACrefauthors}%
\unskip\
\newblock
\APACrefYearMonthDay{2022}{{\APACmonth{09}}}{}.
\newblock
{\BBOQ}\APACrefatitle {Kwai e {TikTok} viralizam fake news e até recomendam
  busca} {Kwai e {TikTok} viralizam fake news e até recomendam busca}.{\BBCQ}
\newblock
\APACjournalVolNumPages{Folha de São Paulo}{}{}{}.
\newblock
\begin{APACrefURL}
  \url{https://www1.folha.uol.com.br/poder/2022/10/kwai-e-tiktok-viralizam-videos-de-fake-news-de-fraude-eleitoral-e-ate-recomendam-material.shtml}
  \end{APACrefURL}
\PrintBackRefs{\CurrentBib}

\bibitem [\protect \citeauthoryear {%
Garimella%
\ \BBA {} Tyson%
}{%
Garimella%
\ \BBA {} Tyson%
}{%
{\protect \APACyear {2018}}%
}]{%
garimella2018whatsapp}
\APACinsertmetastar {%
garimella2018whatsapp}%
\begin{APACrefauthors}%
Garimella, K.%
\BCBT {}\ \BBA {} Tyson, G.%
\end{APACrefauthors}%
\unskip\
\newblock
\APACrefYearMonthDay{2018}{Jun.}{}.
\newblock
{\BBOQ}\APACrefatitle {{WhatApp} Doc? A First Look at {WhatsApp} Public Group
  Data} {{WhatApp} doc? a first look at {WhatsApp} public group data}.{\BBCQ}
\newblock
\APACjournalVolNumPages{Proceedings of the International AAAI Conference on Web
  and Social Media}{12}{1}{}.
\newblock
\begin{APACrefURL}
  \url{https://ojs.aaai.org/index.php/ICWSM/article/view/14989}
  \end{APACrefURL}
\newblock
\begin{APACrefDOI} \doi{10.1609/icwsm.v12i1.14989} \end{APACrefDOI}
\PrintBackRefs{\CurrentBib}

\bibitem [\protect \citeauthoryear {%
Gava%
}{%
Gava%
}{%
{\protect \APACyear {2022}}%
}]{%
gava_estudo_2022}
\APACinsertmetastar {%
gava_estudo_2022}%
\begin{APACrefauthors}%
Gava, M.%
\end{APACrefauthors}%
\unskip\
\newblock
\APACrefYearMonthDay{2022}{}{}.
\newblock
\APACrefbtitle {Estudo reúne dados sobre o uso de redes sociais no {Brasil}.}
  {Estudo reúne dados sobre o uso de redes sociais no {Brasil}.}
\newblock
\begin{APACrefURL}
  \url{https://www.capterra.com.br/blog/3007/uso-redes-sociais}
  \end{APACrefURL}
\PrintBackRefs{\CurrentBib}

\bibitem [\protect \citeauthoryear {%
Gibson%
}{%
Gibson%
}{%
{\protect \APACyear {2014}}%
}]{%
gibson2014theory}
\APACinsertmetastar {%
gibson2014theory}%
\begin{APACrefauthors}%
Gibson, J\BPBI J.%
\end{APACrefauthors}%
\unskip\
\newblock
\APACrefYearMonthDay{2014}{}{}.
\newblock
{\BBOQ}\APACrefatitle {The Theory of Affordances:(1979)} {The theory of
  affordances:(1979)}.{\BBCQ}
\newblock
\BIn{} \APACrefbtitle {The people, place, and space reader} {The people, place,
  and space reader}\ (\BPGS\ 56--60).
\newblock
\APACaddressPublisher{}{Routledge}.
\PrintBackRefs{\CurrentBib}

\bibitem [\protect \citeauthoryear {%
Ha%
, Andreu~Perez%
\BCBL {}\ \BBA {} Ray%
}{%
Ha%
\ \protect \BOthers {.}}{%
{\protect \APACyear {2021}}%
}]{%
ha_mapping_2021}
\APACinsertmetastar {%
ha_mapping_2021}%
\begin{APACrefauthors}%
Ha, L.%
, Andreu~Perez, L.%
\BCBL {}\ \BBA {} Ray, R.%
\end{APACrefauthors}%
\unskip\
\newblock
\APACrefYearMonthDay{2021}{{\APACmonth{02}}}{}.
\newblock
{\BBOQ}\APACrefatitle {Mapping {Recent} {Development} in {Scholarship} on
  {Fake} {News} and {Misinformation}, 2008 to 2017: {Disciplinary}
  {Contribution}, {Topics}, and {Impact}} {Mapping {Recent} {Development} in
  {Scholarship} on {Fake} {News} and {Misinformation}, 2008 to 2017:
  {Disciplinary} {Contribution}, {Topics}, and {Impact}}.{\BBCQ}
\newblock
\APACjournalVolNumPages{American Behavioral Scientist}{65}{2}{290--315}.
\newblock
\begin{APACrefURL} [{2023-09-01}]\url{https://doi.org/10.1177/0002764219869402}
  \end{APACrefURL}
\newblock
\APACrefnote{Publisher: SAGE Publications Inc}
\newblock
\begin{APACrefDOI} \doi{10.1177/0002764219869402} \end{APACrefDOI}
\PrintBackRefs{\CurrentBib}

\bibitem [\protect \citeauthoryear {%
Hecht%
\ \BBA {} Gergle%
}{%
Hecht%
\ \BBA {} Gergle%
}{%
{\protect \APACyear {2010}}%
}]{%
hecht2010tower}
\APACinsertmetastar {%
hecht2010tower}%
\begin{APACrefauthors}%
Hecht, B.%
\BCBT {}\ \BBA {} Gergle, D.%
\end{APACrefauthors}%
\unskip\
\newblock
\APACrefYearMonthDay{2010}{}{}.
\newblock
{\BBOQ}\APACrefatitle {The Tower of Babel Meets Web 2.0: User-Generated Content
  and Its Applications in a Multilingual Context} {The tower of babel meets web
  2.0: User-generated content and its applications in a multilingual
  context}.{\BBCQ}
\newblock
\BIn{} \APACrefbtitle {Proceedings of the SIGCHI Conference on Human Factors in
  Computing Systems} {Proceedings of the sigchi conference on human factors in
  computing systems}\ (\BPG~291–300).
\newblock
\APACaddressPublisher{New York, NY, USA}{Association for Computing Machinery}.
\newblock
\begin{APACrefURL} \url{https://doi.org/10.1145/1753326.1753370}
  \end{APACrefURL}
\newblock
\begin{APACrefDOI} \doi{10.1145/1753326.1753370} \end{APACrefDOI}
\PrintBackRefs{\CurrentBib}

\bibitem [\protect \citeauthoryear {%
Hernon%
}{%
Hernon%
}{%
{\protect \APACyear {1995}}%
}]{%
hernon1995disinformation}
\APACinsertmetastar {%
hernon1995disinformation}%
\begin{APACrefauthors}%
Hernon, P.%
\end{APACrefauthors}%
\unskip\
\newblock
\APACrefYearMonthDay{1995}{}{}.
\newblock
{\BBOQ}\APACrefatitle {Disinformation and misinformation through the internet:
  Findings of an exploratory study} {Disinformation and misinformation through
  the internet: Findings of an exploratory study}.{\BBCQ}
\newblock
\APACjournalVolNumPages{Government Information Quarterly}{12}{2}{133-139}.
\newblock
\begin{APACrefURL}
  \url{https://www.sciencedirect.com/science/article/pii/0740624X95900527}
  \end{APACrefURL}
\newblock
\begin{APACrefDOI} \doi{https://doi.org/10.1016/0740-624X(95)90052-7}
  \end{APACrefDOI}
\PrintBackRefs{\CurrentBib}

\bibitem [\protect \citeauthoryear {%
Jerit%
\ \BBA {} Zhao%
}{%
Jerit%
\ \BBA {} Zhao%
}{%
{\protect \APACyear {2020}}%
}]{%
jerit2020political}
\APACinsertmetastar {%
jerit2020political}%
\begin{APACrefauthors}%
Jerit, J.%
\BCBT {}\ \BBA {} Zhao, Y.%
\end{APACrefauthors}%
\unskip\
\newblock
\APACrefYearMonthDay{2020}{}{}.
\newblock
{\BBOQ}\APACrefatitle {Political Misinformation} {Political
  misinformation}.{\BBCQ}
\newblock
\APACjournalVolNumPages{Annual Review of Political Science}{23}{1}{77-94}.
\newblock
\begin{APACrefURL} \url{https://doi.org/10.1146/annurev-polisci-050718-032814}
  \end{APACrefURL}
\newblock
\begin{APACrefDOI} \doi{10.1146/annurev-polisci-050718-032814} \end{APACrefDOI}
\PrintBackRefs{\CurrentBib}

\bibitem [\protect \citeauthoryear {%
Johansson%
\ \protect \BOthers {.}}{%
Johansson%
\ \protect \BOthers {.}}{%
{\protect \APACyear {2022}}%
}]{%
johansson2022combat}
\APACinsertmetastar {%
johansson2022combat}%
\begin{APACrefauthors}%
Johansson, P.%
, Enock, F.%
, Hale, S.%
, Vidgen, B.%
, Bereskin, C.%
, Margetts, H.%
\BCBL {}\ \BBA {} Bright, J.%
\end{APACrefauthors}%
\unskip\
\newblock
\APACrefYearMonthDay{2022}{}{}.
\newblock
{\BBOQ}\APACrefatitle {How can we combat online misinformation? A systematic
  overview of current interventions and their efficacy} {How can we combat
  online misinformation? a systematic overview of current interventions and
  their efficacy}.{\BBCQ}
\newblock
\APACjournalVolNumPages{arXiv}{}{}{}.
\newblock
\begin{APACrefURL} \url{https://arxiv.org/abs/2212.11864} \end{APACrefURL}
\PrintBackRefs{\CurrentBib}

\bibitem [\protect \citeauthoryear {%
Jungherr%
}{%
Jungherr%
}{%
{\protect \APACyear {2018}}%
}]{%
jungherr2018normalizing}
\APACinsertmetastar {%
jungherr2018normalizing}%
\begin{APACrefauthors}%
Jungherr, A.%
\end{APACrefauthors}%
\unskip\
\newblock
\APACrefYear{2018}.
\newblock
\APACrefbtitle {Normalizing digital trace data} {Normalizing digital trace
  data}.
\newblock
\APACaddressPublisher{}{New York: Routledge}.
\PrintBackRefs{\CurrentBib}

\bibitem [\protect \citeauthoryear {%
Kazemi%
, Garimella%
, Gaffney%
\BCBL {}\ \BBA {} Hale%
}{%
Kazemi%
\ \protect \BOthers {.}}{%
{\protect \APACyear {2021}}%
{\protect \APACexlab {{\protect \BCnt {1}}}}}]{%
kazemi_claim_2021}
\APACinsertmetastar {%
kazemi_claim_2021}%
\begin{APACrefauthors}%
Kazemi, A.%
, Garimella, K.%
, Gaffney, D.%
\BCBL {}\ \BBA {} Hale, S.%
\end{APACrefauthors}%
\unskip\
\newblock
\APACrefYearMonthDay{2021{\protect \BCnt {1}}}{{\APACmonth{08}}}{}.
\newblock
{\BBOQ}\APACrefatitle {Claim {Matching} {Beyond} {English} to {Scale} {Global}
  {Fact}-{Checking}} {Claim {Matching} {Beyond} {English} to {Scale} {Global}
  {Fact}-{Checking}}.{\BBCQ}
\newblock
\BIn{} \APACrefbtitle {Proceedings of the 59th {Annual} {Meeting} of the
  {Association} for {Computational} {Linguistics} and the 11th {International}
  {Joint} {Conference} on {Natural} {Language} {Processing} ({Volume} 1: {Long}
  {Papers})} {Proceedings of the 59th {Annual} {Meeting} of the {Association}
  for {Computational} {Linguistics} and the 11th {International} {Joint}
  {Conference} on {Natural} {Language} {Processing} ({Volume} 1: {Long}
  {Papers})}\ (\BPGS\ 4504--4517).
\newblock
\APACaddressPublisher{}{Association for Computational Linguistics}.
\newblock
\begin{APACrefURL} \url{https://aclanthology.org/2021.acl-long.347}
  \end{APACrefURL}
\newblock
\begin{APACrefDOI} \doi{10.18653/v1/2021.acl-long.347} \end{APACrefDOI}
\PrintBackRefs{\CurrentBib}

\bibitem [\protect \citeauthoryear {%
Kazemi%
, Garimella%
, Gaffney%
\BCBL {}\ \BBA {} Hale%
}{%
Kazemi%
\ \protect \BOthers {.}}{%
{\protect \APACyear {2021}}%
{\protect \APACexlab {{\protect \BCnt {2}}}}}]{%
kazemi2021claim}
\APACinsertmetastar {%
kazemi2021claim}%
\begin{APACrefauthors}%
Kazemi, A.%
, Garimella, K.%
, Gaffney, D.%
\BCBL {}\ \BBA {} Hale, S.%
\end{APACrefauthors}%
\unskip\
\newblock
\APACrefYearMonthDay{2021{\protect \BCnt {2}}}{{\APACmonth{08}}}{}.
\newblock
{\BBOQ}\APACrefatitle {Claim Matching Beyond {E}nglish to Scale Global
  Fact-Checking} {Claim matching beyond {E}nglish to scale global
  fact-checking}.{\BBCQ}
\newblock
\BIn{} \APACrefbtitle {Proceedings of the 59th Annual Meeting of the
  Association for Computational Linguistics and the 11th International Joint
  Conference on Natural Language Processing (Volume 1: Long Papers)}
  {Proceedings of the 59th annual meeting of the association for computational
  linguistics and the 11th international joint conference on natural language
  processing (volume 1: Long papers)}\ (\BPGS\ 4504--4517).
\newblock
\APACaddressPublisher{Online}{Association for Computational Linguistics}.
\newblock
\begin{APACrefURL} \url{https://aclanthology.org/2021.acl-long.347}
  \end{APACrefURL}
\newblock
\begin{APACrefDOI} \doi{10.18653/v1/2021.acl-long.347} \end{APACrefDOI}
\PrintBackRefs{\CurrentBib}

\bibitem [\protect \citeauthoryear {%
Kazemi%
, Garimella%
, Shahi%
, Gaffney%
\BCBL {}\ \BBA {} Hale%
}{%
Kazemi%
\ \protect \BOthers {.}}{%
{\protect \APACyear {2022}}%
}]{%
kazemi2022tiplines}
\APACinsertmetastar {%
kazemi2022tiplines}%
\begin{APACrefauthors}%
Kazemi, A.%
, Garimella, K.%
, Shahi, G\BPBI K.%
, Gaffney, D.%
\BCBL {}\ \BBA {} Hale, S\BPBI A.%
\end{APACrefauthors}%
\unskip\
\newblock
\APACrefYearMonthDay{2022}{}{}.
\newblock
{\BBOQ}\APACrefatitle {Tiplines to uncover misinformation on encrypted
  platforms: A case study of the 2019 {Indian} general election on {WhatsApp}}
  {Tiplines to uncover misinformation on encrypted platforms: A case study of
  the 2019 {Indian} general election on {WhatsApp}}.{\BBCQ}
\newblock
\APACjournalVolNumPages{Harvard Kennedy School (HKS) Misinformation
  Review}{}{}{}.
\newblock
\begin{APACrefURL} \url{https://doi.org/10.37016/mr-2020-91} \end{APACrefURL}
\PrintBackRefs{\CurrentBib}

\bibitem [\protect \citeauthoryear {%
Kemp%
}{%
Kemp%
}{%
{\protect \APACyear {2022}}%
}]{%
kemp_digital_2022}
\APACinsertmetastar {%
kemp_digital_2022}%
\begin{APACrefauthors}%
Kemp, S.%
\end{APACrefauthors}%
\unskip\
\newblock
\APACrefYearMonthDay{2022}{}{}.
\newblock
\APACrefbtitle {Digital 2022: {Brazil}} {Digital 2022: {Brazil}}\
  \APACbVolEdTR{}{\BTR{}}.
\newblock
\APACaddressInstitution{}{Data Reportal}.
\newblock
\begin{APACrefURL} \url{https://datareportal.com/reports/digital-2022-brazil}
  \end{APACrefURL}
\PrintBackRefs{\CurrentBib}

\bibitem [\protect \citeauthoryear {%
Kischinhevsky%
\ \protect \BOthers {.}}{%
Kischinhevsky%
\ \protect \BOthers {.}}{%
{\protect \APACyear {2020}}%
}]{%
kischinhevsky_whatsapp_2020}
\APACinsertmetastar {%
kischinhevsky_whatsapp_2020}%
\begin{APACrefauthors}%
Kischinhevsky, M.%
, Vieira, I\BPBI M.%
, Santos, J\BPBI G\BPBI B\BPBI d.%
, Chagas, V.%
, Freitas, M\BPBI d\BPBI A.%
\BCBL {}\ \BBA {} Aldé, A.%
\end{APACrefauthors}%
\unskip\
\newblock
\APACrefYearMonthDay{2020}{{\APACmonth{10}}}{}.
\newblock
{\BBOQ}\APACrefatitle {{WhatsApp} audios and the remediation of radio:
  {Disinformation} in {Brazilian} 2018 presidential election} {{WhatsApp}
  audios and the remediation of radio: {Disinformation} in {Brazilian} 2018
  presidential election}.{\BBCQ}
\newblock
\APACjournalVolNumPages{Radio Journal:International Studies in Broadcast \&
  Audio Media}{18}{2}{139--158}.
\newblock
\begin{APACrefURL}
  \url{https://intellectdiscover.com/content/journals/10.1386/rjao_00021_1}
  \end{APACrefURL}
\newblock
\begin{APACrefDOI} \doi{10.1386/rjao_00021_1} \end{APACrefDOI}
\PrintBackRefs{\CurrentBib}

\bibitem [\protect \citeauthoryear {%
Lamb%
, King%
\BCBL {}\ \BBA {} Kling%
}{%
Lamb%
\ \protect \BOthers {.}}{%
{\protect \APACyear {2003}}%
}]{%
lamb2003informational}
\APACinsertmetastar {%
lamb2003informational}%
\begin{APACrefauthors}%
Lamb, R.%
, King, J\BPBI L.%
\BCBL {}\ \BBA {} Kling, R.%
\end{APACrefauthors}%
\unskip\
\newblock
\APACrefYearMonthDay{2003}{}{}.
\newblock
{\BBOQ}\APACrefatitle {Informational environments: Organizational contexts of
  online information use} {Informational environments: Organizational contexts
  of online information use}.{\BBCQ}
\newblock
\APACjournalVolNumPages{Journal of the American Society for Information Science
  and Technology}{54}{2}{97--114}.
\PrintBackRefs{\CurrentBib}

\bibitem [\protect \citeauthoryear {%
\APACcitebtitle {{L}anguage {L}og: {S}nowclones: lexicographical dating to the
  second --- itre.cis.upenn.edu}}{%
\APACcitebtitle {{L}anguage {L}og: {S}nowclones: lexicographical dating to the
  second --- itre.cis.upenn.edu}}{%
{\protect \APACyear {{\protect \bibnodate {}}}}%
}]{%
upennLanguageLog}
\APACinsertmetastar {%
upennLanguageLog}%
\APACrefbtitle {{L}anguage {L}og: {S}nowclones: lexicographical dating to the
  second --- itre.cis.upenn.edu.} {{L}anguage {L}og: {S}nowclones:
  lexicographical dating to the second --- itre.cis.upenn.edu.}
\newblock
\APACrefYearMonthDay{{\protect \bibnodate {}}}{}{}.
\newblock
\APAChowpublished
  {\url{http://itre.cis.upenn.edu/~myl/languagelog/archives/000350.html}}.
\newblock
\APACrefnote{[Accessed 01-Feb-2023]}
\PrintBackRefs{\CurrentBib}

\bibitem [\protect \citeauthoryear {%
Lelo%
}{%
Lelo%
}{%
{\protect \APACyear {2022}}%
}]{%
thales2022rise}
\APACinsertmetastar {%
thales2022rise}%
\begin{APACrefauthors}%
Lelo, T.%
\end{APACrefauthors}%
\unskip\
\newblock
\APACrefYearMonthDay{2022}{}{}.
\newblock
{\BBOQ}\APACrefatitle {The Rise of the Brazilian Fact-checking Movement:
  Between Economic Sustainability and Editorial Independence} {The rise of the
  brazilian fact-checking movement: Between economic sustainability and
  editorial independence}.{\BBCQ}
\newblock
\APACjournalVolNumPages{Journalism Studies}{23}{9}{1077-1095}.
\newblock
\begin{APACrefURL} \url{https://doi.org/10.1080/1461670X.2022.2069588}
  \end{APACrefURL}
\newblock
\begin{APACrefDOI} \doi{10.1080/1461670X.2022.2069588} \end{APACrefDOI}
\PrintBackRefs{\CurrentBib}

\bibitem [\protect \citeauthoryear {%
Litt%
}{%
Litt%
}{%
{\protect \APACyear {2012}}%
}]{%
litt2012imagined}
\APACinsertmetastar {%
litt2012imagined}%
\begin{APACrefauthors}%
Litt, E.%
\end{APACrefauthors}%
\unskip\
\newblock
\APACrefYearMonthDay{2012}{}{}.
\newblock
{\BBOQ}\APACrefatitle {Knock, Knock. Who's There? The Imagined Audience}
  {Knock, knock. who's there? the imagined audience}.{\BBCQ}
\newblock
\APACjournalVolNumPages{Journal of Broadcasting \& Electronic
  Media}{56}{3}{330-345}.
\newblock
\begin{APACrefURL} \url{https://doi.org/10.1080/08838151.2012.705195}
  \end{APACrefURL}
\newblock
\begin{APACrefDOI} \doi{10.1080/08838151.2012.705195} \end{APACrefDOI}
\PrintBackRefs{\CurrentBib}

\bibitem [\protect \citeauthoryear {%
Machado%
, Kira%
, Narayanan%
, Kollanyi%
\BCBL {}\ \BBA {} Howard%
}{%
Machado%
\ \protect \BOthers {.}}{%
{\protect \APACyear {2019}}%
}]{%
machado_study_2019}
\APACinsertmetastar {%
machado_study_2019}%
\begin{APACrefauthors}%
Machado, C.%
, Kira, B.%
, Narayanan, V.%
, Kollanyi, B.%
\BCBL {}\ \BBA {} Howard, P.%
\end{APACrefauthors}%
\unskip\
\newblock
\APACrefYearMonthDay{2019}{{\APACmonth{05}}}{}.
\newblock
{\BBOQ}\APACrefatitle {A {Study} of {Misinformation} in {WhatsApp} groups with
  a focus on the {Brazilian} {Presidential} {Elections}.} {A {Study} of
  {Misinformation} in {WhatsApp} groups with a focus on the {Brazilian}
  {Presidential} {Elections}.}{\BBCQ}
\newblock
\BIn{} \APACrefbtitle {Companion {Proceedings} of {The} 2019 {World} {Wide}
  {Web} {Conference}} {Companion {Proceedings} of {The} 2019 {World} {Wide}
  {Web} {Conference}}\ (\BPGS\ 1013--1019).
\newblock
\APACaddressPublisher{San Francisco USA}{ACM}.
\newblock
\begin{APACrefURL}
  [{2023-01-22}]\url{https://dl.acm.org/doi/10.1145/3308560.3316738}
  \end{APACrefURL}
\newblock
\begin{APACrefDOI} \doi{10.1145/3308560.3316738} \end{APACrefDOI}
\PrintBackRefs{\CurrentBib}

\bibitem [\protect \citeauthoryear {%
Margetts%
, John%
, Hale%
\BCBL {}\ \BBA {} Yasseri%
}{%
Margetts%
\ \protect \BOthers {.}}{%
{\protect \APACyear {2015}}%
}]{%
margetts2015political}
\APACinsertmetastar {%
margetts2015political}%
\begin{APACrefauthors}%
Margetts, H.%
, John, P.%
, Hale, S.%
\BCBL {}\ \BBA {} Yasseri, T.%
\end{APACrefauthors}%
\unskip\
\newblock
\APACrefYear{2015}.
\newblock
\APACrefbtitle {Political Turbulence: How Social Media Shape Collective Action}
  {Political turbulence: How social media shape collective action}.
\PrintBackRefs{\CurrentBib}

\bibitem [\protect \citeauthoryear {%
Metzger%
, Flanagin%
, Mena%
, Jiang%
\BCBL {}\ \BBA {} Wilson%
}{%
Metzger%
\ \protect \BOthers {.}}{%
{\protect \APACyear {2021}}%
}]{%
metzger2021from}
\APACinsertmetastar {%
metzger2021from}%
\begin{APACrefauthors}%
Metzger, M.%
, Flanagin, A.%
, Mena, P.%
, Jiang, S.%
\BCBL {}\ \BBA {} Wilson, C.%
\end{APACrefauthors}%
\unskip\
\newblock
\APACrefYearMonthDay{2021}{}{}.
\newblock
{\BBOQ}\APACrefatitle {From Dark to Light: The Many Shades of Sharing
  Misinformation Online} {From dark to light: The many shades of sharing
  misinformation online}.{\BBCQ}
\newblock
\APACjournalVolNumPages{Media and Communication}{9}{1}{134--143}.
\newblock
\begin{APACrefURL}
  \url{https://www.cogitatiopress.com/mediaandcommunication/article/view/3409}
  \end{APACrefURL}
\newblock
\begin{APACrefDOI} \doi{10.17645/mac.v9i1.3409} \end{APACrefDOI}
\PrintBackRefs{\CurrentBib}

\bibitem [\protect \citeauthoryear {%
Newman%
, Fletcher%
, Robertson%
, Eddy%
\BCBL {}\ \BBA {} Nielsen%
}{%
Newman%
\ \protect \BOthers {.}}{%
{\protect \APACyear {2022}}%
}]{%
newman_digital_2022}
\APACinsertmetastar {%
newman_digital_2022}%
\begin{APACrefauthors}%
Newman, N.%
, Fletcher, R.%
, Robertson, C\BPBI T.%
, Eddy, K.%
\BCBL {}\ \BBA {} Nielsen, R\BPBI K.%
\end{APACrefauthors}%
\unskip\
\newblock
\APACrefYearMonthDay{2022}{}{}.
\newblock
\APACrefbtitle {Digital {News} {Report} 2022} {Digital {News} {Report} 2022}\
  \APACbVolEdTR{}{\BTR{}}.
\newblock
\APACaddressInstitution{}{Reuters Institute \& University of Oxford}.
\newblock
\begin{APACrefURL}
  \url{https://reutersinstitute.politics.ox.ac.uk/digital-news-report/2022}
  \end{APACrefURL}
\PrintBackRefs{\CurrentBib}

\bibitem [\protect \citeauthoryear {%
Norman%
}{%
Norman%
}{%
{\protect \APACyear {2013}}%
}]{%
norman2013design}
\APACinsertmetastar {%
norman2013design}%
\begin{APACrefauthors}%
Norman, D.%
\end{APACrefauthors}%
\unskip\
\newblock
\APACrefYear{2013}.
\newblock
\APACrefbtitle {The design of everyday things: Revised and expanded edition}
  {The design of everyday things: Revised and expanded edition}.
\newblock
\APACaddressPublisher{}{Basic books}.
\PrintBackRefs{\CurrentBib}

\bibitem [\protect \citeauthoryear {%
Oliveira%
\ \BBA {} Carmo%
}{%
Oliveira%
\ \BBA {} Carmo%
}{%
{\protect \APACyear {2022}}%
}]{%
oliveira_memorando_2022}
\APACinsertmetastar {%
oliveira_memorando_2022}%
\begin{APACrefauthors}%
Oliveira, R\BPBI M\BPBI d.%
\BCBT {}\ \BBA {} Carmo, L\BPBI L\BPBI d.%
\end{APACrefauthors}%
\unskip\
\newblock
\APACrefYearMonthDay{2022}{}{}.
\newblock
\APACrefbtitle {Memorando de {Entendimento} - {TSE} n. 6/2022.} {Memorando de
  {Entendimento} - {TSE} n. 6/2022.}
\newblock
\APACaddressPublisher{}{Tribunal Superior Eleitoral}.
\newblock
\begin{APACrefURL}
  [{2023-01-14}]\url{https://www.tse.jus.br/comunicacao/noticias/arquivos/assinatura-de-acordos-plataformas-digitais/memorando-tse-e-twitter/@@download/file/MoU
  TSE_Twitter.pdf} \end{APACrefURL}
\PrintBackRefs{\CurrentBib}

\bibitem [\protect \citeauthoryear {%
Oliveira%
\ \BBA {} Zarife%
}{%
Oliveira%
\ \BBA {} Zarife%
}{%
{\protect \APACyear {2021}}%
}]{%
oliveira_memorando_2021}
\APACinsertmetastar {%
oliveira_memorando_2021}%
\begin{APACrefauthors}%
Oliveira, R\BPBI M\BPBI d.%
\BCBT {}\ \BBA {} Zarife, F\BPBI O.%
\end{APACrefauthors}%
\unskip\
\newblock
\APACrefYearMonthDay{2021}{}{}.
\newblock
\APACrefbtitle {Memorando de {Entendimento} - {TSE} n. 23/2021.} {Memorando de
  {Entendimento} - {TSE} n. 23/2021.}
\newblock
\APACaddressPublisher{}{Tribunal Superior Eleitoral}.
\newblock
\begin{APACrefURL}
  [{2023-01-14}]\url{https://www.tse.jus.br/comunicacao/noticias/arquivos/assinatura-de-acordos-plataformas-digitais/memorando-tse-e-kwai/@@download/file/MoU
  Kwai_20222.pdf} \end{APACrefURL}
\PrintBackRefs{\CurrentBib}

\bibitem [\protect \citeauthoryear {%
Osmundsen%
, Bor%
, Vahlstrup%
, Bechmann%
\BCBL {}\ \BBA {} Petersen%
}{%
Osmundsen%
\ \protect \BOthers {.}}{%
{\protect \APACyear {2021}}%
}]{%
osmundsen2021partisan}
\APACinsertmetastar {%
osmundsen2021partisan}%
\begin{APACrefauthors}%
Osmundsen, M.%
, Bor, A.%
, Vahlstrup, P\BPBI B.%
, Bechmann, A.%
\BCBL {}\ \BBA {} Petersen, M\BPBI B.%
\end{APACrefauthors}%
\unskip\
\newblock
\APACrefYearMonthDay{2021}{}{}.
\newblock
{\BBOQ}\APACrefatitle {Partisan Polarization Is the Primary Psychological
  Motivation behind Political Fake News Sharing on Twitter} {Partisan
  polarization is the primary psychological motivation behind political fake
  news sharing on twitter}.{\BBCQ}
\newblock
\APACjournalVolNumPages{American Political Science Review}{115}{3}{999–1015}.
\newblock
\begin{APACrefDOI} \doi{10.1017/S0003055421000290} \end{APACrefDOI}
\PrintBackRefs{\CurrentBib}

\bibitem [\protect \citeauthoryear {%
Paiva%
}{%
Paiva%
}{%
{\protect \APACyear {2020}}%
}]{%
whatsapp99}
\APACinsertmetastar {%
whatsapp99}%
\begin{APACrefauthors}%
Paiva, F.%
\end{APACrefauthors}%
\unskip\
\newblock
\APACrefYearMonthDay{2020}{}{}.
\newblock
\APACrefbtitle {{WhatsApp} alcança presença recorde em 99%
  Brasil - Mobile Time.} {{WhatsApp} alcança presença recorde em 99%
  smartphones no brasil - mobile time.}
\newblock
\APAChowpublished
  {https://www.mobiletime.com.br/noticias/27/02/2020/whatsapp-alcanca-presenca-recorde-em-99-dos-smartphones-no-brasil/}.
\newblock
\APACrefnote{[Accessed 01-Feb-2023]}
\PrintBackRefs{\CurrentBib}

\bibitem [\protect \citeauthoryear {%
Paiva%
}{%
Paiva%
}{%
{\protect \APACyear {2022}}%
}]{%
paiva_panorama_2022}
\APACinsertmetastar {%
paiva_panorama_2022}%
\begin{APACrefauthors}%
Paiva, F.%
\end{APACrefauthors}%
\unskip\
\newblock
\APACrefYearMonthDay{2022}{}{}.
\newblock
\APACrefbtitle {Panorama {Mobile} {Time}/{Opinion} {Box}} {Panorama {Mobile}
  {Time}/{Opinion} {Box}}\ \APACbVolEdTR{}{\BTR{}}.
\newblock
\begin{APACrefURL} \url{https://www.mobiletime.com.br/pesquisas/}
  \end{APACrefURL}
\PrintBackRefs{\CurrentBib}

\bibitem [\protect \citeauthoryear {%
Panciera%
, Halfaker%
\BCBL {}\ \BBA {} Terveen%
}{%
Panciera%
\ \protect \BOthers {.}}{%
{\protect \APACyear {2009}}%
}]{%
panciera2009wikipedians}
\APACinsertmetastar {%
panciera2009wikipedians}%
\begin{APACrefauthors}%
Panciera, K.%
, Halfaker, A.%
\BCBL {}\ \BBA {} Terveen, L.%
\end{APACrefauthors}%
\unskip\
\newblock
\APACrefYearMonthDay{2009}{}{}.
\newblock
{\BBOQ}\APACrefatitle {Wikipedians Are Born, Not Made: A Study of Power Editors
  on Wikipedia} {Wikipedians are born, not made: A study of power editors on
  wikipedia}.{\BBCQ}
\newblock
\BIn{} \APACrefbtitle {Proceedings of the 2009 ACM International Conference on
  Supporting Group Work} {Proceedings of the 2009 acm international conference
  on supporting group work}\ (\BPG~51–60).
\newblock
\APACaddressPublisher{New York, NY, USA}{Association for Computing Machinery}.
\newblock
\begin{APACrefURL} \url{https://doi.org/10.1145/1531674.1531682}
  \end{APACrefURL}
\newblock
\begin{APACrefDOI} \doi{10.1145/1531674.1531682} \end{APACrefDOI}
\PrintBackRefs{\CurrentBib}

\bibitem [\protect \citeauthoryear {%
Pantazi%
, Hale%
\BCBL {}\ \BBA {} Klein%
}{%
Pantazi%
\ \protect \BOthers {.}}{%
{\protect \APACyear {2021}}%
}]{%
pantazi2021social}
\APACinsertmetastar {%
pantazi2021social}%
\begin{APACrefauthors}%
Pantazi, M.%
, Hale, S.%
\BCBL {}\ \BBA {} Klein, O.%
\end{APACrefauthors}%
\unskip\
\newblock
\APACrefYearMonthDay{2021}{}{}.
\newblock
{\BBOQ}\APACrefatitle {Social and Cognitive Aspects of the Vulnerability to
  Political Misinformation} {Social and cognitive aspects of the vulnerability
  to political misinformation}.{\BBCQ}
\newblock
\APACjournalVolNumPages{Political Psychology}{42}{S1}{267-304}.
\newblock
\begin{APACrefURL}
  \url{https://onlinelibrary.wiley.com/doi/abs/10.1111/pops.12797}
  \end{APACrefURL}
\newblock
\begin{APACrefDOI} \doi{https://doi.org/10.1111/pops.12797} \end{APACrefDOI}
\PrintBackRefs{\CurrentBib}

\bibitem [\protect \citeauthoryear {%
Pennycook%
\ \protect \BOthers {.}}{%
Pennycook%
\ \protect \BOthers {.}}{%
{\protect \APACyear {2021}}%
}]{%
pennycook2021shifting}
\APACinsertmetastar {%
pennycook2021shifting}%
\begin{APACrefauthors}%
Pennycook, G.%
, Epstein, Z.%
, Mosleh, M.%
, Arechar, A\BPBI A.%
, Eckles, D.%
\BCBL {}\ \BBA {} Rand, D\BPBI G.%
\end{APACrefauthors}%
\unskip\
\newblock
\APACrefYearMonthDay{2021}{}{}.
\newblock
{\BBOQ}\APACrefatitle {Shifting attention to accuracy can reduce misinformation
  online} {Shifting attention to accuracy can reduce misinformation
  online}.{\BBCQ}
\newblock
\APACjournalVolNumPages{Nature}{592}{7855}{590--595}.
\newblock
\begin{APACrefURL} \url{https://doi.org/10.1038/s41586-021-03344-2}
  \end{APACrefURL}
\PrintBackRefs{\CurrentBib}

\bibitem [\protect \citeauthoryear {%
Pereira%
, Lemos%
, Adami%
\BCBL {}\ \BBA {} Carvalho%
}{%
Pereira%
\ \protect \BOthers {.}}{%
{\protect \APACyear {2019}}%
}]{%
pereira_compatibility_2019}
\APACinsertmetastar {%
pereira_compatibility_2019}%
\begin{APACrefauthors}%
Pereira, C\BPBI M\BPBI d\BPBI S.%
, Lemos, R.%
, Adami, M\BPBI P.%
\BCBL {}\ \BBA {} Carvalho, F\BPBI M\BPBI d.%
\end{APACrefauthors}%
\unskip\
\newblock
\APACrefYearMonthDay{2019}{{\APACmonth{08}}}{}.
\newblock
{\BBOQ}\APACrefatitle {Compatibility of zero-rating offers with {Brazilian} net
  neutrality rules} {Compatibility of zero-rating offers with {Brazilian} net
  neutrality rules}.{\BBCQ}
\newblock
\APACjournalVolNumPages{Revista Direito GV}{15}{}{}.
\newblock
\begin{APACrefURL}
  \url{http://www.scielo.br/j/rdgv/a/y7VMVvXQPMktCXP3fBgPHwm/abstract/?lang=en}
  \end{APACrefURL}
\newblock
\begin{APACrefDOI} \doi{10.1590/2317-6172201919} \end{APACrefDOI}
\PrintBackRefs{\CurrentBib}

\bibitem [\protect \citeauthoryear {%
Recuero%
}{%
Recuero%
}{%
{\protect \APACyear {2020}}%
}]{%
recuero_fraudenasurnas_2020}
\APACinsertmetastar {%
recuero_fraudenasurnas_2020}%
\begin{APACrefauthors}%
Recuero, R.%
\end{APACrefauthors}%
\unskip\
\newblock
\APACrefYearMonthDay{2020}{{\APACmonth{08}}}{}.
\newblock
{\BBOQ}\APACrefatitle {\#{FraudenasUrnas}: estratégias discursivas de
  desinformação no {Twitter} nas eleições 2018} {\#{FraudenasUrnas}:
  estratégias discursivas de desinformação no {Twitter} nas eleições
  2018}.{\BBCQ}
\newblock
\APACjournalVolNumPages{Revista Brasileira de Linguística
  Aplicada}{20}{}{383--406}.
\newblock
\begin{APACrefURL}
  \url{http://www.scielo.br/j/rbla/a/vKnghPRMJxbypBVRLYN3YTB/abstract/?lang=pt}
  \end{APACrefURL}
\newblock
\begin{APACrefDOI} \doi{10.1590/1984-6398202014635} \end{APACrefDOI}
\PrintBackRefs{\CurrentBib}

\bibitem [\protect \citeauthoryear {%
Reimers%
\ \BBA {} Gurevych%
}{%
Reimers%
\ \BBA {} Gurevych%
}{%
{\protect \APACyear {2019}}%
}]{%
reimers-gurevych-2019-sentence}
\APACinsertmetastar {%
reimers-gurevych-2019-sentence}%
\begin{APACrefauthors}%
Reimers, N.%
\BCBT {}\ \BBA {} Gurevych, I.%
\end{APACrefauthors}%
\unskip\
\newblock
\APACrefYearMonthDay{2019}{{\APACmonth{11}}}{}.
\newblock
{\BBOQ}\APACrefatitle {Sentence-{BERT}: Sentence Embeddings using {S}iamese
  {BERT}-Networks} {Sentence-{BERT}: Sentence embeddings using {S}iamese
  {BERT}-networks}.{\BBCQ}
\newblock
\BIn{} \APACrefbtitle {Proceedings of the 2019 Conference on Empirical Methods
  in Natural Language Processing and the 9th International Joint Conference on
  Natural Language Processing (EMNLP-IJCNLP)} {Proceedings of the 2019
  conference on empirical methods in natural language processing and the 9th
  international joint conference on natural language processing
  (emnlp-ijcnlp)}\ (\BPGS\ 3982--3992).
\newblock
\APACaddressPublisher{Hong Kong, China}{Association for Computational
  Linguistics}.
\newblock
\begin{APACrefURL} \url{https://aclanthology.org/D19-1410} \end{APACrefURL}
\newblock
\begin{APACrefDOI} \doi{10.18653/v1/D19-1410} \end{APACrefDOI}
\PrintBackRefs{\CurrentBib}

\bibitem [\protect \citeauthoryear {%
J.~Reis%
, Melo%
, Garimella%
\BCBL {}\ \BBA {} Benevenuto%
}{%
J.~Reis%
\ \protect \BOthers {.}}{%
{\protect \APACyear {2020}}%
}]{%
reis2020whatsapp}
\APACinsertmetastar {%
reis2020whatsapp}%
\begin{APACrefauthors}%
Reis, J.%
, Melo, P.%
, Garimella, K.%
\BCBL {}\ \BBA {} Benevenuto, F.%
\end{APACrefauthors}%
\unskip\
\newblock
\APACrefYearMonthDay{2020}{}{}.
\newblock
{\BBOQ}\APACrefatitle {Can {WhatsApp} benefit from debunked fact-checked
  stories to reduce misinformation?} {Can {WhatsApp} benefit from debunked
  fact-checked stories to reduce misinformation?}{\BBCQ}
\newblock
\APACjournalVolNumPages{Harvard Kennedy School (HKS) Misinformation
  Review}{}{}{}.
\newblock
\begin{APACrefURL} \url{https://doi.org/10.37016/mr-2020-035} \end{APACrefURL}
\PrintBackRefs{\CurrentBib}

\bibitem [\protect \citeauthoryear {%
J\BPBI C\BPBI S.~Reis%
, Melo%
, Garimella%
, Almeida%
\BCBL {}\ \protect \BOthers {.}}{%
J\BPBI C\BPBI S.~Reis%
, Melo%
, Garimella%
, Almeida%
\BCBL {}\ \protect \BOthers {.}}{%
{\protect \APACyear {2020}}%
}]{%
reis_dataset_2020}
\APACinsertmetastar {%
reis_dataset_2020}%
\begin{APACrefauthors}%
Reis, J\BPBI C\BPBI S.%
, Melo, P.%
, Garimella, K.%
, Almeida, J\BPBI M.%
, Eckles, D.%
\BCBL {}\ \BBA {} Benevenuto, F.%
\end{APACrefauthors}%
\unskip\
\newblock
\APACrefYearMonthDay{2020}{{\APACmonth{05}}}{}.
\newblock
{\BBOQ}\APACrefatitle {A {Dataset} of {Fact}-{Checked} {Images} {Shared} on
  {WhatsApp} {During} the {Brazilian} and {Indian} {Elections}} {A {Dataset} of
  {Fact}-{Checked} {Images} {Shared} on {WhatsApp} {During} the {Brazilian} and
  {Indian} {Elections}}.{\BBCQ}
\newblock
\APACjournalVolNumPages{Proceedings of the International AAAI Conference on Web
  and Social Media}{14}{}{903--908}.
\newblock
\begin{APACrefURL}
  [{2023-01-22}]\url{https://ojs.aaai.org/index.php/ICWSM/article/view/7356}
  \end{APACrefURL}
\newblock
\begin{APACrefDOI} \doi{10.1609/icwsm.v14i1.7356} \end{APACrefDOI}
\PrintBackRefs{\CurrentBib}

\bibitem [\protect \citeauthoryear {%
J\BPBI C\BPBI S.~Reis%
, Melo%
, Garimella%
\BCBL {}\ \BBA {} Benevenuto%
}{%
J\BPBI C\BPBI S.~Reis%
, Melo%
, Garimella%
\BCBL {}\ \BBA {} Benevenuto%
}{%
{\protect \APACyear {2020}}%
}]{%
reis_can_2020}
\APACinsertmetastar {%
reis_can_2020}%
\begin{APACrefauthors}%
Reis, J\BPBI C\BPBI S.%
, Melo, P\BPBI d\BPBI F.%
, Garimella, K.%
\BCBL {}\ \BBA {} Benevenuto, F.%
\end{APACrefauthors}%
\unskip\
\newblock
\APACrefYearMonthDay{2020}{{\APACmonth{08}}}{}.
\newblock
\APACrefbtitle {Can {WhatsApp} {Benefit} from {Debunked} {Fact}-{Checked}
  {Stories} to {Reduce} {Misinformation}?} {Can {WhatsApp} {Benefit} from
  {Debunked} {Fact}-{Checked} {Stories} to {Reduce} {Misinformation}?}
\newblock
\begin{APACrefURL} \url{http://arxiv.org/abs/2006.02471} \end{APACrefURL}
\newblock
\APACrefnote{arXiv:2006.02471 [cs]}
\newblock
\begin{APACrefDOI} \doi{10.48550/arXiv.2006.02471} \end{APACrefDOI}
\PrintBackRefs{\CurrentBib}

\bibitem [\protect \citeauthoryear {%
Resende%
, Melo%
, C.~S.~Reis%
\BCBL {}\ \protect \BOthers {.}}{%
Resende%
, Melo%
, C.~S.~Reis%
\BCBL {}\ \protect \BOthers {.}}{%
{\protect \APACyear {2019}}%
}]{%
resende_analyzing_2019}
\APACinsertmetastar {%
resende_analyzing_2019}%
\begin{APACrefauthors}%
Resende, G.%
, Melo, P.%
, C.~S.~Reis, J.%
, Vasconcelos, M.%
, Almeida, J\BPBI M.%
\BCBL {}\ \BBA {} Benevenuto, F.%
\end{APACrefauthors}%
\unskip\
\newblock
\APACrefYearMonthDay{2019}{{\APACmonth{06}}}{}.
\newblock
{\BBOQ}\APACrefatitle {Analyzing {Textual} ({Mis}){Information} {Shared} in
  {WhatsApp} {Groups}} {Analyzing {Textual} ({Mis}){Information} {Shared} in
  {WhatsApp} {Groups}}.{\BBCQ}
\newblock
\BIn{} \APACrefbtitle {Proceedings of the 10th {ACM} {Conference} on {Web}
  {Science}} {Proceedings of the 10th {ACM} {Conference} on {Web} {Science}}\
  (\BPGS\ 225--234).
\newblock
\APACaddressPublisher{New York, NY, USA}{Association for Computing Machinery}.
\newblock
\begin{APACrefURL} \url{https://doi.org/10.1145/3292522.3326029}
  \end{APACrefURL}
\newblock
\begin{APACrefDOI} \doi{10.1145/3292522.3326029} \end{APACrefDOI}
\PrintBackRefs{\CurrentBib}

\bibitem [\protect \citeauthoryear {%
Resende%
, Melo%
, Sousa%
\BCBL {}\ \protect \BOthers {.}}{%
Resende%
, Melo%
, Sousa%
\BCBL {}\ \protect \BOthers {.}}{%
{\protect \APACyear {2019}}%
}]{%
resende_misinformation_2019}
\APACinsertmetastar {%
resende_misinformation_2019}%
\begin{APACrefauthors}%
Resende, G.%
, Melo, P.%
, Sousa, H.%
, Messias, J.%
, Vasconcelos, M.%
, Almeida, J.%
\BCBL {}\ \BBA {} Benevenuto, F.%
\end{APACrefauthors}%
\unskip\
\newblock
\APACrefYearMonthDay{2019}{{\APACmonth{05}}}{}.
\newblock
{\BBOQ}\APACrefatitle {({Mis}){Information} {Dissemination} in {WhatsApp}:
  {Gathering}, {Analyzing} and {Countermeasures}} {({Mis}){Information}
  {Dissemination} in {WhatsApp}: {Gathering}, {Analyzing} and
  {Countermeasures}}.{\BBCQ}
\newblock
\BIn{} \APACrefbtitle {The {World} {Wide} {Web} {Conference}} {The {World}
  {Wide} {Web} {Conference}}\ (\BPGS\ 818--828).
\newblock
\APACaddressPublisher{New York, NY, USA}{Association for Computing Machinery}.
\newblock
\begin{APACrefURL} [{2023-01-24}]\url{https://doi.org/10.1145/3308558.3313688}
  \end{APACrefURL}
\newblock
\begin{APACrefDOI} \doi{10.1145/3308558.3313688} \end{APACrefDOI}
\PrintBackRefs{\CurrentBib}

\bibitem [\protect \citeauthoryear {%
Ressa%
}{%
Ressa%
}{%
{\protect \APACyear {2023}}%
}]{%
ressanobel}
\APACinsertmetastar {%
ressanobel}%
\begin{APACrefauthors}%
Ressa, M.%
\end{APACrefauthors}%
\unskip\
\newblock
\APACrefYearMonthDay{2023}{}{}.
\newblock
\APACrefbtitle {Nobel Prize lecture. NobelPrize.org. Nobel Prize Outreach {AB}
  2023.} {Nobel prize lecture. nobelprize.org. nobel prize outreach {AB} 2023.}
\newblock
\APAChowpublished
  {https://www.nobelprize.org/prizes/peace/2021/ressa/lecture/}.
\newblock
\APACrefnote{[Accessed 01-Feb-2023]}
\PrintBackRefs{\CurrentBib}

\bibitem [\protect \citeauthoryear {%
Rossini%
, Mont’Alverne%
\BCBL {}\ \BBA {} Kalogeropoulos%
}{%
Rossini%
\ \protect \BOthers {.}}{%
{\protect \APACyear {2023}}%
}]{%
rossini_explaining_2023}
\APACinsertmetastar {%
rossini_explaining_2023}%
\begin{APACrefauthors}%
Rossini, P.%
, Mont’Alverne, C.%
\BCBL {}\ \BBA {} Kalogeropoulos, A.%
\end{APACrefauthors}%
\unskip\
\newblock
\APACrefYearMonthDay{2023}{{\APACmonth{05}}}{}.
\newblock
{\BBOQ}\APACrefatitle {Explaining beliefs in electoral misinformation in the
  2022 {Brazilian} election: {The} role of ideology, political trust, social
  media, and messaging apps} {Explaining beliefs in electoral misinformation in
  the 2022 {Brazilian} election: {The} role of ideology, political trust,
  social media, and messaging apps}.{\BBCQ}
\newblock
\APACjournalVolNumPages{Harvard Kennedy School Misinformation Review}{}{}{}.
\newblock
\begin{APACrefURL}
  \url{https://misinforeview.hks.harvard.edu/article/explaining-beliefs-in-electoral-misinformation-in-the-2022-brazilian-election-the-role-of-ideology-political-trust-social-media-and-messaging-apps/}
  \end{APACrefURL}
\newblock
\begin{APACrefDOI} \doi{10.37016/mr-2020-115} \end{APACrefDOI}
\PrintBackRefs{\CurrentBib}

\bibitem [\protect \citeauthoryear {%
Ruths%
\ \BBA {} Pfeffer%
}{%
Ruths%
\ \BBA {} Pfeffer%
}{%
{\protect \APACyear {2014}}%
}]{%
ruths2014social}
\APACinsertmetastar {%
ruths2014social}%
\begin{APACrefauthors}%
Ruths, D.%
\BCBT {}\ \BBA {} Pfeffer, J.%
\end{APACrefauthors}%
\unskip\
\newblock
\APACrefYearMonthDay{2014}{}{}.
\newblock
{\BBOQ}\APACrefatitle {Social media for large studies of behavior} {Social
  media for large studies of behavior}.{\BBCQ}
\newblock
\APACjournalVolNumPages{Science}{346}{6213}{1063--1064}.
\PrintBackRefs{\CurrentBib}

\bibitem [\protect \citeauthoryear {%
Scofield%
\ \BBA {} Fonseca%
}{%
Scofield%
\ \BBA {} Fonseca%
}{%
{\protect \APACyear {2022}}%
}]{%
scofield_tiktok_2022}
\APACinsertmetastar {%
scofield_tiktok_2022}%
\begin{APACrefauthors}%
Scofield, L.%
\BCBT {}\ \BBA {} Fonseca, N.%
\end{APACrefauthors}%
\unskip\
\newblock
\APACrefYearMonthDay{2022}{{\APACmonth{12}}}{}.
\newblock
\APACrefbtitle {{TikTok} e {Kwai} levam desinformação sobre urnas e {Forças}
  {Armadas} ao {WhatsApp}.} {{TikTok} e {Kwai} levam desinformação sobre
  urnas e {Forças} {Armadas} ao {WhatsApp}.}
\newblock
\begin{APACrefURL}
  \url{https://apublica.org/sentinela/2022/09/tiktok-e-kwai-levam-desinformacao-sobre-urnas-e-forcas-armadas-ao-whatsapp/}
  \end{APACrefURL}
\PrintBackRefs{\CurrentBib}

\bibitem [\protect \citeauthoryear {%
Shaar%
, Babulkov%
, Da~San~Martino%
\BCBL {}\ \BBA {} Nakov%
}{%
Shaar%
\ \protect \BOthers {.}}{%
{\protect \APACyear {2020}}%
}]{%
shaar_that_2020}
\APACinsertmetastar {%
shaar_that_2020}%
\begin{APACrefauthors}%
Shaar, S.%
, Babulkov, N.%
, Da~San~Martino, G.%
\BCBL {}\ \BBA {} Nakov, P.%
\end{APACrefauthors}%
\unskip\
\newblock
\APACrefYearMonthDay{2020}{{\APACmonth{07}}}{}.
\newblock
{\BBOQ}\APACrefatitle {That is a {Known} {Lie}: {Detecting} {Previously}
  {Fact}-{Checked} {Claims}} {That is a {Known} {Lie}: {Detecting} {Previously}
  {Fact}-{Checked} {Claims}}.{\BBCQ}
\newblock
\BIn{} \APACrefbtitle {Proceedings of the 58th {Annual} {Meeting} of the
  {Association} for {Computational} {Linguistics}} {Proceedings of the 58th
  {Annual} {Meeting} of the {Association} for {Computational} {Linguistics}}\
  (\BPGS\ 3607--3618).
\newblock
\APACaddressPublisher{}{Association for Computational Linguistics}.
\newblock
\begin{APACrefURL} \url{https://aclanthology.org/2020.acl-main.332}
  \end{APACrefURL}
\newblock
\begin{APACrefDOI} \doi{10.18653/v1/2020.acl-main.332} \end{APACrefDOI}
\PrintBackRefs{\CurrentBib}

\bibitem [\protect \citeauthoryear {%
Tarouco%
}{%
Tarouco%
}{%
{\protect \APACyear {2023}}%
}]{%
tarouco_brazilian_2023}
\APACinsertmetastar {%
tarouco_brazilian_2023}%
\begin{APACrefauthors}%
Tarouco, G.%
\end{APACrefauthors}%
\unskip\
\newblock
\APACrefYearMonthDay{2023}{}{}.
\newblock
{\BBOQ}\APACrefatitle {Brazilian 2022 general elections: process, results, and
  implications} {Brazilian 2022 general elections: process, results, and
  implications}.{\BBCQ}
\newblock
\APACjournalVolNumPages{Revista Uruguaya de Ciencia Política}{32}{1}{}.
\newblock
\begin{APACrefURL}
  \url{http://www.scielo.edu.uy/scielo.php?script=sci_abstract&pid=S1688-499X2023000100153&lng=es&nrm=iso&tlng=en}
  \end{APACrefURL}
\newblock
\begin{APACrefDOI} \doi{10.26851/rucp.32.1.7} \end{APACrefDOI}
\PrintBackRefs{\CurrentBib}

\bibitem [\protect \citeauthoryear {%
Tomaz%
\ \BBA {} Tomaz%
}{%
Tomaz%
\ \BBA {} Tomaz%
}{%
{\protect \APACyear {2020}}%
}]{%
tomaz_brazilian_2020}
\APACinsertmetastar {%
tomaz_brazilian_2020}%
\begin{APACrefauthors}%
Tomaz, R\BPBI M.%
\BCBT {}\ \BBA {} Tomaz, J\BPBI M\BPBI T.%
\end{APACrefauthors}%
\unskip\
\newblock
\APACrefYearMonthDay{2020}{{\APACmonth{01}}}{}.
\newblock
{\BBOQ}\APACrefatitle {The {Brazilian} {Presidential} {Election} of 2018 and
  the relationship between technology and democracy in {Latin} {America}} {The
  {Brazilian} {Presidential} {Election} of 2018 and the relationship between
  technology and democracy in {Latin} {America}}.{\BBCQ}
\newblock
\APACjournalVolNumPages{Journal of Information, Communication and Ethics in
  Society}{18}{4}{}.
\newblock
\begin{APACrefURL} \url{https://doi.org/10.1108/JICES-12-2019-0134}
  \end{APACrefURL}
\newblock
\APACrefnote{Publisher: Emerald Publishing Limited}
\newblock
\begin{APACrefDOI} \doi{10.1108/JICES-12-2019-0134} \end{APACrefDOI}
\PrintBackRefs{\CurrentBib}

\bibitem [\protect \citeauthoryear {%
TSE%
}{%
TSE%
}{%
{\protect \APACyear {2022}}%
}]{%
tse_eleicoes_2022}
\APACinsertmetastar {%
tse_eleicoes_2022}%
\begin{APACrefauthors}%
TSE.%
\end{APACrefauthors}%
\unskip\
\newblock
\APACrefYearMonthDay{2022}{{\APACmonth{01}}}{}.
\newblock
\APACrefbtitle {Eleições 2022: {TSE} e {WhatsApp} discutem medidas para
  enfrentamento da desinformação.} {Eleições 2022: {TSE} e {WhatsApp}
  discutem medidas para enfrentamento da desinformação.}
\newblock
\begin{APACrefURL}
  [{2023-01-25}]\url{https://www.tse.jus.br/comunicacao/noticias/2022/Janeiro/eleicoes-2022-tse-e-whatsapp-discutem-medidas-para-enfrentamento-da-desinformacao}
  \end{APACrefURL}
\PrintBackRefs{\CurrentBib}

\end{thebibliography}
\end{document}